\documentclass{emulateapj}

\usepackage{graphics,psfig}

\newcommand{\mum}{$\mu$m}
\newcommand{\ha}{H$\alpha$}
\newcommand{\hb}{H$\beta$}

\newcommand{\hii}{\ion{H}{2}}

\newcommand{\oiii}{[\ion{O}{3}]}
\newcommand{\oii}{[\ion{O}{2}]}
\newcommand{\caiih}{\ion{Ca}{2}\ H}
\newcommand{\caiik}{\ion{Ca}{2}\ K}
\newcommand{\msun}{M$_\odot$}
\newcommand{\lsun}{L$_\odot$}
\newcommand{\zsun}{Z$_\odot$}

\newcommand{\ebv}{$E_{\rm B-\rm V}$}

\begin{document}
\title{Star Formation History and Extinction in the central kpc of 
M\,82-like Starbursts} 

\author{
Y.D. Mayya\altaffilmark{1}, 
A. Bressan\altaffilmark{2,3},
M. Rodr\'{\i}guez\altaffilmark{1}, 
J.R. Valdes\altaffilmark{1,2}, and 
M. Chavez\altaffilmark{1,4}
}

\altaffiltext{1}
{Instituto Nacional de Astrof\'{\i}sica, Optica y Electr\'onica,
Luis Enrique Erro 1, Tonantzintla, Apdo Postal 51 y 216, C.P. 72840,
Puebla, M\'exico}
\altaffiltext{2}
{Osservatorio Astronomico di Padova, Vicolo dell'Osservatorio 535122, Padova, Italy}
\altaffiltext{3}
{Scuola Internazionale Superiore di Studi Avanzati,
Via Beirut 4, I-34014 Trieste, Italy}
\altaffiltext{4}
{Currently at LPL and Steward Observatory, University of Arizona, Tucson, AZ 85719} 
\email{ydm@inaoep.mx, bressan@pd.astro.it, mrodri@inaoep.mx, jvaldes@inaoep.mx, mchavez@inaoep.mx}

\slugcomment{Scheduled to appear in ApJ Jan 1, 2004}
\shortauthors{Mayya, Bressan, Rodriguez, Valdes, Chavez}
\shorttitle{Nearby Starburst Galaxies}

\begin{abstract}
We report on the star formation histories and extinction in
the central kpc region of a sample of starburst galaxies that have similar 
far infrared (FIR), 10$\mu$m and K-band luminosities as those of the 
archetype starburst M\,82. 
Our study is based on new optical spectra and previously published
K-band photometric data, both sampling the same area around the nucleus.
Model starburst spectra were synthesized as a combination of stellar 
populations of  distinct ages formed over the Hubble time, and were 
fitted to the observed optical spectra and K-band flux. 
The model is able to reproduce simultaneously the equivalent widths of 
emission and absorption lines, the continuum fluxes between 3500--7000\,\AA,
the K-band and the FIR flux. We require a minimum of 
3 populations --- (1) a young population of age $\le$8~Myr, with its
corresponding nebular emission,
(2) an intermediate-age population (age $<$ 500 Myr), and 
(3) an old population that forms part of the underlying disk or/and 
bulge population. 
The birthrate parameter, which is defined as the ratio of the current star 
formation rate to the average past rate, is found to be in the range 1 to 12.
The contribution of the old population to the K-band luminosity
depends on the birthrate parameter and remains above 60\% in the
majority of the sample galaxies. 
Even in the blue band, the intermediate age and old populations
contribute more than 40\% of the total flux in all the cases.
A relatively high contribution from the old stars to the
K-band nuclear flux is also apparent from the strength of the 4000\,\AA\ break 
and the \caiik\ line. 
The extinction of the old population is found to be around half of that of
the young population. 
The contribution to the continuum from the relatively old stars has the 
effect of diluting the emission equivalent widths below the values expected
for young bursts. The mean dilution factors are found to be 
5 and 3 for the \ha\ and \hb\ lines respectively.

\end{abstract}


\section{Introduction}

Starburst galaxies harbor thousands of massive stars, that are
responsible for their typical nebular spectrum and high far infrared (FIR) 
emission (Weedman et al. 1981). Massive  stars are short-lived, typically 
less than 10 million years (Myr), consequently, most of the derived 
properties of the starburst regions pertain to the last generation of
massive stars. On the other hand, starburst galaxies have potential to
form stars for periods much longer than this.
For example M\,82, the prototype of a starburst, has sufficient amount of 
gas in order to sustain star formation for a few hundreds of million years
(Walter, Weiss \& Scoville 2002).
But, does the star formation really continue for periods much longer than the 
typical age of massive stars? By their very nature, studies based on 
emission lines are unable to answer this question. We need techniques that use
tracers of lower mass stars, which live for several tens or
even a few hundreds of million years. 

The optical spectrum of these older stellar systems are
characterized by prominent Balmer absorption lines. Indeed, Kim et al. (1995)
found such absorption lines to be common in starburst spectra.
In recent years, there have been attempts to use the information contained
in the absorption lines to derive the properties of one or more assumed 
older bursts in individual galaxies, e.g. NGC\,7714 (Lancon et al. 2001) 
and NGC\,7679 (Gu et al. 2001). Stellar populations of a few hundreds 
of million years have been inferred in these studies.
However, these stellar features are not being routinely used to derive 
properties such as the age and mass of the older stellar populations of a
representative sample of galaxies. 
This is probably because of difficulties in analyzing the
absorption features in the presence of nebular emission.
Recently, Poggianti, Bressan \& Franceschini (2001) have developed a 
technique based on synthetic spectra that allows investigation of 
many previous bursts simultaneously.
Their technique is capable of handling Balmer emission lines even in the
presence of an underlying absorption line, and vice versa, and hence is ideal 
for an analysis of the complete star formation history of starburst regions. 

The star formation history of a starburst nucleus depends on the spatial 
scale under investigation. Observations of nearby starburst galaxies 
with the Hubble Space Telescope have resolved starburst nuclei into 
several tens, or in some cases, a few hundreds of compact clusters 
known as Super Star Clusters (SSCs) (O'Connell et al. 1995).
SSCs show a considerable range in their colors, suggesting a spread in 
their ages. Typically SSCs are found in a region of a few hundred parsecs 
within the starburst nucleus. In M\,82, regions with strong Balmer 
absorption lines are found as far as 1~kpc from the currently active 
starburst site (O'Connell \&  Mangano 1978). If we were to place M\,82 at 
the distance of Arp\,220, an Ultra Luminous Far Infrared Galaxy, 
all the regions discussed above would be inside an aperture of 3\arcsec.
Thus ground-based spectra of distant starburst nuclei would include not
only the present burst, but also the outlying older bursts, if any.
Therefore, we need to study the star formation history of nearby
starburst nuclei over a spatial scale of $\sim$1\,kpc in order to compare
the results with those of distant galaxies.

Star-forming regions are always associated with dust, which has the effect
of reddening the observed spectrum. The amount of reddening is traditionally
determined by comparing the observed Balmer emission line ratios with 
theoretical values (Osterbrock 1989), and making use of
the Galactic reddening curve (Cardelli  et al. 1989).
An alternative method to find the reddening is to compare the observed
ultraviolet continuum slope of the star-forming regions 
with that expected for young starbursts from synthetic models
 (Fanelli, O'Connel \& Thuan 1988).
Using the above two techniques, it is now well established that
the reddening suffered by the emission lines in starburst regions
is significantly higher than the reddening suffered by the continuum 
(Fanelli et al. 1988; Calzetti, Kinney \& Storchi-Bergmann 1994). 
The difference in the reddening is understood to be due to the 
escape of stars from the parent molecular clouds as they get older
(Calzetti et al. 1994; Silva et al. 1998). 
Calzetti et al. (1994) also found that the mean attenuation curve towards the 
starburst regions is more gray as compared to that of the Milky Way.
Charlot \& Fall (2000) investigated these issues
in detail and concluded that a key ingredient that is responsible for a 
differential reddening between lines and continuum, and a gray reddening 
curve, is the finite lifetime of the stellar birth clouds. 
Granato et al. (2001) found that the starburst spectra could be 
modeled by even using the Galactic reddening curve, when each stellar 
population constituting the starburst is allowed to have its own extinction.
It is important to note that the low observed strength of the 2200\,\AA\
ultraviolet bump in starburst spectra could also be produced by the
models of Granato et al.

In this work, we study the star formation histories of the central kpc region
of a sample of starburst galaxies, focusing on determining the masses of 
all the stellar populations that contribute to the optical spectrum. 
The relative contributions of the young and old populations to the K-band 
flux are also determined.
We use the entire optical spectrum, not just a few emission and absorption
lines, so that our observables are sensitive not only to the young burst,
but also to all the previous bursts. The starburst model of 
Poggianti et al. (2001) is used.
The extinction of each population is a free parameter to be evaluated
by the fitting procedure. 
In Sec.~2, we describe
the sample and observations. The starburst model is explained in Sec.~3.
The star formation history of the sample galaxies is discussed in Sec.~4.
Concluding remarks are given in Sec.~5.

\section{Sample, Observations and Analysis}

Devereux (1989) defined a complete sample of {\it nearby starburst galaxies},
that includes all nearby, 16$\le$ D(Mpc) $\le$40
(Hubble constant of 75 km\,s$^{-1}$\,Mpc$^{-1}$),
non-Seyfert 1 galaxies with central 10$\mu$m luminosity $\ge6\times10^8$
\lsun, declination $-35^\circ\le \delta \le +60^\circ$, right ascension 
$22^{\rm h} \le \alpha \le 16^{\rm h}$, and Galactic latitude $b\ge 20^\circ$.
The sample contains 20 galaxies, whose K-band and FIR
luminosities are very similar to those of the prototypical starburst M\,82.
The FIR luminosities, as estimated from 60 and 100\,$\mu$m 
InfraRed Astronomical Satellite (IRAS) fluxes, lie 
between $10^{10}$--$10^{11}$~\lsun, which are intermediate between those
of normal galaxies and Ultra Luminous Far Infrared Galaxies.
We carried out long-slit optical spectrophotometry of 12 of these galaxies 
at the Guillermo Haro Astrophysical Observatory, Cananea, Mexico, 
during two observing runs in 1999 February and December. 
A grating with 150~lines\,mm$^{-1}$ was used, which resulted in a resolution of 
about 10\,\AA, and a spectral coverage  $\approx$3500--6800\,\AA.
The spectral and spatial samplings were 3.2\,\AA\,pix$^{-1}$ and 
0.46\arcsec\,pix$^{-1}$ respectively. Slit widths were typically 
of $1.6^{\prime\prime}$, centered 
on the starburst nucleus. The slit was aligned along the East-West direction 
for the sample galaxies. The objects were observed as close to the meridian 
as possible, with maximum airmass of $\approx1.30$. Two spectra of 30 
minutes each were taken for each galaxy. The instrumental response was 
calibrated by the observation of the standard stars HR\,1544, HR\,5501, 
BD+40 4032, Feige 15 and Feige 34. 

Each frame was bias-corrected and divided by a normalized flat-field
using various tasks in the IRAF package. Two wavelength-calibrated frames of 
the same galaxy were averaged, in the process removing cosmic ray events.
Sky spectra were extracted from the object-free regions of the long-slit 
and were subtracted from each starburst spectrum. The spectra of the sample 
starbursts were extracted over a slit length of $9^{\prime\prime}$,
which corresponds to physical scales of 0.77 to 1.76~kpc. 
The average signal-to-noise ratios of the extracted spectra 
are $\approx$50 and 20 around 5500\,\AA\ and 3700\,\AA\ respectively. 
We estimate that the spectrophotometry is
accurate to better than 10\% over the entire wavelength range covered.
\begin{center}
\begin{deluxetable*}{rlcccrrcccc}
\tabletypesize{\small}
\tablewidth{0pc}
\tablecaption{General properties of the sample starburst nuclei}
\tablehead{
\colhead{Galaxy} &  \colhead{Type}   & \colhead{Distance}  
                 & \colhead{L$_{\rm FIR}$\tablenotemark{b}}
                 &  \colhead{$m_{\rm V}$\tablenotemark{a}}   
                 &  \colhead{\scriptsize FIR/V$_{\rm ap}$}  
                 &  \colhead{\scriptsize FIR/V$_{\rm sp}$}  
                 &  \colhead{F$_{\rm K}$/F$_{\rm V}$}  
                 & \colhead{B$-$V} 
                 & \colhead{A$_{\rm V}$} 
                 & \colhead{Z/\zsun} \\
\colhead{(NGC)}  & \colhead{(RC3)} & \colhead{(Mpc)}  & \colhead{}
                 &  \colhead{(mag)}  
                 &  \colhead{}  
                 &  \colhead{}  
                 &  \colhead{}  & \colhead{(mag)} 
                 &  \colhead{(mag)}  
                 & \colhead{} 
}
\startdata
 470 & Sb    & 30.50 & 10.13 & 14.65 & 16.62 & 37.73 &0.39 & 0.83 &2.31 &3.95 \\
1022 & SBa   & 18.50 & 10.13 & 14.15 & 28.82 &100.84 &0.31 & 0.70 &3.32 &2.91 \\
2273 & SBa   & 28.40 & 10.03 & 13.95 &  6.84 & 16.42 &0.35 & 1.02 &3.74 &0.20 \\
2750 & SABc  & 38.40 & 10.17 & 14.15 &  7.12 & 24.47 &0.19 & 0.39 &2.84 &2.37 \\
2782 & SABa  & 37.30 & 10.39 & 13.92 & 10.84 & 19.19 &0.26 & 0.56 &2.04 &1.49 \\
3504 & SABab & 26.50 & 10.47 & 13.10 & 11.04 & 23.50 &0.24 & 0.60 &3.30 &2.40 \\
4102 & SABb  & 17.00 & 10.44 & 13.00 & 25.00 & 56.75 &0.49 & 1.03 &5.25 &0.43 \\
4194 & IBm   & 39.10 & 10.82 & 13.55 & 18.63 & 44.27 &0.24 & 0.48 &3.02 &1.17 \\
4273 & SBc   & 35.10 & 10.49 & 14.10 & 17.53 & 67.90 &0.21 & 0.66 &2.69 &2.96 \\
4385 & SB0   & 33.50 & 10.01 & 14.45 &  8.77 & 17.82 &0.16 & 0.30 &1.17 &2.40 \\
4818 & SABab & 21.50 & 10.26 & 13.41 & 13.77 & 78.52 &0.55 & 1.05 &4.65 &3.51 \\
4984 & SAB0  & 21.30 & 10.01 & 12.80 &  4.19 & 26.42 &0.59 & 1.00 &3.53 &1.98 \\
\hline
\enddata
\tablenotetext{a}
{V-band magnitude in 9\arcsec\ aperture.} 
\tablenotetext{b}
{Luminosities in log(L/\lsun) units.} 
\label{table1}
\end{deluxetable*}
\end{center}
The general properties of the observed starburst nuclei are presented in 
Table~\ref{table1}. The Hubble types and distances to the galaxies are given 
in columns~2 and 3 respectively. The distances are based
on the recessional velocities in the Nearby Galaxy Catalog (Tully 1988).
Column~4 contains the FIR luminosity  (L$_{\rm FIR}$), which was calculated 
from the 60 and 100 $\mu$m IRAS fluxes (Helou et al. 1988).
IRAS beams, in general, enclose the entire galaxy, and hence a part of the
IRAS flux could originate in regions outside the nucleus. However
recent high resolution mid-infrared images of starburst galaxies obtained
with the Infrared Satellite Observatory (ISO) show 
that most of the IRAS flux originates within a spatial
scale of around 1~kpc (Foerster-Schreiber et al. 2003), which is the 
slit-length used for the extraction of spectra of our sample galaxies.
We estimated the V-band magnitudes in an aperture of $9\arcsec$ ($V_{\rm ap}$) 
using the compilation of multi-aperture photometric data by Prugniel \& 
Heraudeau (1998). These magnitudes are tabulated in Column~5.
V-band magnitudes were also obtained from our slit spectra ($V_{\rm sp}$).
The two magnitudes were measured with the specific purpose of providing two 
approximate limits on the ratio between FIR and V-band luminosity (FIR/V).
The FIR/V ratios using the aperture and slit V-band fluxes are given 
in columns 6 and 7 respectively
(FIR/V = L$_{\rm FIR}$/(L$_{5500}\times 5500$), where L$_{5500}$ is the 
monochromatic luminosity in the $V$-band). For compact sources, the 
real values are expected to be close to FIR/$V_{\rm sp}$, 
whereas for extended sources, they may be as low as FIR/$V_{\rm ap}$.
The K-band flux inside our slit (see section~3.3), normalized to the slit 
V-band flux is given in column~8. 
Column~9 contains the B$-$V color obtained from the extracted spectra.
The visual extinction (A$_{\rm v}$) and the gas metallicity $Z$ are 
given in columns~10 and 11 respectively. 

A$_{\rm v}$ was 
obtained by comparing the ratio of H$\alpha$ to H$\beta$ to its Case B 
value, making use of the Galactic extinction curve (Osterbrock 1989).  
However, the measurement of A$_{\rm v}$ in the presence of underlying 
absorption is not trivial (Rosa-Gonz\'alez, Terlevich \& Terlevich 2002). 
The measured values are somewhat sensitive to the assumed values of the 
equivalent widths of the underlying absorption lines, and this can increase
the uncertainty in $Z$. 
McCall, Rybski, \& Shields (1985) demonstrated that the
stellar absorption correction and the extinction for each 
galaxy could be estimated if the H$\gamma$ line is detected in emission.
They recommended a mean absorption correction of 2\,\AA\ to the equivalent 
widths of the first three lines in Balmer series. However the H$\gamma$ 
emission line is not seen in five of the sample galaxies.
Hence we cannot follow the method used by McCall et al. (1985) uniformly
for all the galaxies.
Alternatively, we 
followed an iterative method to obtain the best estimate of A$_{\rm v}$. 
First, we used the observed spectra with a 2\,\AA\  correction to the
Balmer equivalent widths, to make a preliminary estimation of the
nebular abundances. Then the entire spectral fitting analysis, described in 
detail in section 3.1, was carried out using these abundances.
We then measured the visual extinction on the resulting young population 
model component. This component is not affected by the underlying 
line absorption arising from the intermediate-age populations.
The abundance measurement was refined using this new value of the extinction. 
The strong line method was used to
derive the oxygen abundance from the observed nebular lines using the
calibration of McCall et al. (1985).
The oxygen abundance was converted to metallicity in solar units by
assuming that $12+\log(\rm {O\over{\rm H}})=8.74$ for the Sun (Holweger 2001).

\section{The Theoretical Analysis}
\label{models}

In principle, the analysis of the equivalent width (EW) of the hydrogen 
emission 
lines (e.g. H$\beta$) provides information on the age of the most recent burst 
of star formation (Leitherer et al. 1999), and the Balmer decrement gives the 
extinction values.  However, as discussed in the introduction, a typical slit
may enclose several stellar populations of different ages and a complex 
distribution of dust. Thus in practice, the EW method is heavily affected 
by the uncertainty concerning the contributions of intermediate age and old 
populations to the continuum (and even line absorption) below the emission 
lines. 

On the other hand, the observed spectral energy distributions (SEDs) contain 
much more information than what can be derived from a simple application 
of the Balmer decrement method and equivalent width analysis. The physical 
insight can be even higher if one uses the IRAS fluxes and K band photometry. 
The FIR emission provides a strong constraint on the current star formation 
rate (SFR), though the difference in aperture sizes between the IRAS data 
and our spectra may constitute a serious problem. As for the nuclear K-band 
luminosity, there is a debate as to whether it originates from the old 
population or is provided by red supergiant stars associated with the recent 
star formation history (see Section~4.2).
In order to obtain a complete picture of star formation and extinction
in starburst galaxies,
we have analyzed the SEDs by means of the population synthesis technique.

\subsection{The Starburst Model}

The technique adopted to model the observed spectra is an improved 
version of that described by Poggianti et al. (2001). The star formation 
history over the Hubble time is broken up into several episodes. 
Each episode is described by a burst of star formation, whose rate is constant 
over a certain period of time. Evolution of the spectra within a single
episode is assumed to be negligible and hence each episode is
represented by a SED of a simple stellar population (SSP). 
A SSP is characterized by an age and metallicity.
The ages of SSPs are suitably selected to provide the main spectral 
characteristics seen in starburst spectra (emission and absorption 
features and the overall continuum) 
and are combined together with appropriate intensity and extinction. 

We used nine SSPs, eight of them formed over the last 1~Gyr, while 
the 9$^{\rm th}$ one represents the underlying old population. 
Ages of the first eight SSPs are 2, 7, 10, 50, 100, 300, 500 and 1000~Myr.
The duration of each star formation episode is chosen in such a way that
the SED remains representative of all the ages those are included
by the episode.
The resulting durations are 4, 4, 7, 60, 125, 200, 350 and 1250~Myr 
respectively.
The only place where we would be using these durations is in
the conversion between the stellar mass and the star formation rate.
Henceforth, episodes of star formation corresponding to the first three
SSPs will be referred to as the ``young population'', those 
for the next four SSPs as the ``intermediate age population'', 
and the populations older than $\sim$1 Gyr as the ``old population''.
The ``old population'' is meant to represent the sum of the bulge and disk 
components that enter the slit area.

For each SSP we adopt a uniform screen attenuation, using 
the standard extinction law for the diffuse medium in our Galaxy 
(R$_{\rm V}=$A$_{\rm V}$/E$_{\rm B-\rm V}=3.1$, Cardelli et al. 1989). 
While a more complex picture of 
the extinction cannot be excluded, it was comprehensively shown by 
Poggianti et al. (2001) that the characteristics of the emerging spectrum 
require a significant 
amount of {\sl foreground} dust (screen model). Indeed in the case of a 
uniform mixture of dust and stars, increasing the obscuration does not 
yield a corresponding increase in the {\em reddening} of the spectrum: the
latter saturates to a value E$_{\rm B-\rm V}\sim0.18$, which is too low to 
account for the observed emission line ratios (see also Calzetti et al. 1994). 
The extinction value A$_{\rm V}$ is allowed to vary for the different 
stellar populations and the extinguished spectral energy distributions of 
all the single generations are added up to give the total integrated spectrum.

Allowing the extinction of each population to vary results in a degeneracy 
between extinction and star formation rate, in such a way that there can be 
populations with very high SFR, but completely extinguished, thus 
contributing insignificantly to the total SED. If the extinguished 
population is old, then it will not even contribute to the FIR emission.
This degeneracy can increase somewhat arbitrarily the 
predicted mass to light ratio of the system. In order to avoid such 
extinguished populations, we automatically tested and excluded from the
solution those intermediate and old populations which contribute 
less than 1\% to the total flux at all wavelengths. 
Although dusty young populations may not contribute significantly to the 
optical bands, they are important contributors 
to the FIR emission, and hence the populations younger than 10~Myr are 
retained irrespective of their contribution to the total flux at any 
wavelength. In this way, the algorithm naturally selects not more 
than four SSPs.
This is consistent with the time resolution of a general population synthesis
technique, when one takes into account observational errors and 
uncertainties of the models.

Our SSPs have a Salpeter-like initial mass function 
(IMF) --- N(M)dM $\propto M^{-2.35}$dM --- with masses between 0.15 
and 120\,\msun. The stellar spectral atlas is that of Jacoby, Hunter \& 
Christian (1984), extended in the UV and near infrared with Kurucz (1993)  
models as described in Bressan, Granato \& Silva (1998).

The young population spectrum includes the nebular spectrum, which is
calculated by means of case B photoionization models computed with 
CLOUDY (Ferland 1996) with the following parameters: 
mass of the ionizing cluster 10$^5$\,\msun, 
electron number density n=100\,cm$^{-3}$, 
inner radius 15 pc and metallicity rescaled to that of the SSP. Since our 
main interest is to model the hydrogen spectral features (emission and 
absorption lines), changing these parameters would not affect our conclusions 
significantly.

The gas metallicity of the majority of our sample of starburst galaxies, where 
the metallicity could be reliably measured, is super-solar.
Possible exceptions are NGC\,2273 and NGC\,4102. The former galaxy 
is known to have a weak Seyfert component (Devereux 1989), which is 
possibly contaminating the \oiii$\lambda$5007 line,
thus resulting in an under-estimation of its metallicity.
NGC\,4102 is the most extinguished galaxy of the sample, and its
low metallicity could be due to a possible over-correction to
the extinction of the \oii$\lambda3627$ line.
Thus in general, the young populations are metal rich. 
However the metallicity of the old stars must be lower than that; in
the Galactic bulge the average value is around half solar 
(Sadler, Rich \& Terndrup 1996). Thus, except for NGC\,2273, we have 
adopted Z=2.5\,\zsun\ for the intermediate-young populations and  Z=0.5
\zsun\ for the old population. For NGC\,2273, we have adopted Z=\zsun\  
for the young stellar populations. The results do not depend critically
on the adopted metallicity.

To evaluate the best-fit model we minimize a merit function (MF) which was
constructed from the EWs of H$\delta$, H$\beta$ and H$\alpha$ and the relative 
intensities of the continuum flux in 12 almost featureless narrow 
spectral windows (3580--3680\,\AA, 3845--3880\,\AA, 3900--3920\,\AA,
4000--4060\,\AA, 4140--4300\,\AA, 4400--4600\,\AA, 4700--4820\,\AA, 
5080--5180\,\AA, 5400--5700\,\AA, 5950--6200\,\AA, 6350--6450\,\AA $\,$ and 
6630--6660\,\AA). A MF is obtained from the differences between modeled and
observed values at the selected spectral windows/features, following
the definition:
\begin{equation}\label{MF}
{\rm MF}=\sum_{i=1}^{n}\left(\frac{M_i-O_i}{E_i}\cdot W_i\right)^2
\end{equation}
where $M_i$ and $O_i$ are respectively the model and observed fluxes (or EWs),
and $E_i$ and $W_i$ are the corresponding errors and weights. The errors 
in the spectral window were estimated as the rms fluctuations of the observed 
flux in the corresponding wavelength interval, while
errors of EWs are expressed as a percentage value
above a minimum value of  0.1\,\AA,  0.1\,\AA\  and 1\,\AA,  respectively
for H$\delta$, H$\beta$ and H$\alpha$. 
The minimization procedure applies the algorithm of Adaptive Simulated
Annealing (Ingber 1989) for a continuum multi-dimensional parameter space,
to a merit function. It is important to stress that
the algorithm of Adaptive Simulated Annealing is able to avoid local 
minima and provides convergence to the global minimum. However, the 
uncertainties of the synthesis technique combined with the observational 
errors do not allow us to distinguish sub-populations within the same 
broad age range: young, intermediate-age, and old. For the same reason 
equally good fits can be obtained by making use of SSPs of reasonably 
different metal content (i.e. solar or half-solar for the intermediate 
age and old populations).

Additional constraints that may also be considered, such as the observed ratio
between the FIR to V-band luminosity and the K to V-band luminosity, 
are discussed below.

\subsection{The Far Infrared Emission}

All objects have published IRAS fluxes at 12, 25, 60 and 100 \mum\ 
(Devereux 1987)
so that in principle we might constrain the total light that is absorbed 
in the model by comparison with the observed total infrared luminosity.
By analogy with the observational quantity FIR/V, defined in the previous 
section (see Table \ref{table1}), we compute a theoretical quantity, FIR/V, 
representing the ratio between the absorbed light and the quantity 
F$_\lambda$(5550\AA)$\times$5550 (erg\,s$^{-1}$). The absorbed light is  
easily obtained as the difference between the attenuated and the intrinsic 
light of the stellar populations that provide the best fit.

However it is not easy to compare with the corresponding observational 
quantity, because only a fraction of the radiation emitted in the FIR 
could be sampled by our slit, whose aperture is smaller than the IRAS beam and 
covers only the central region ($\simeq$1 kpc) of the galaxy. 
As discussed in the previous section, aperture problems limit the
use of the FIR/V values.
We thus decided to consider the FIR/V ratio only as an output
of the model, to be compared with the values provided in Table \ref{table1}.
It is worth noting that a model that fits the optical data provides
only a lower limit to the expected FIR emission (for the assumed 
extinction law), because of the possible presence of heavily attenuated 
young populations contributing to the FIR, but not to the optical emission.
For extinction curves that are flatter than the galactic curve,
the model FIR luminosities would be smaller than the values reported in this
study.

\subsection{The Near Infrared Emission}

For all our program galaxies, Devereux (1989) has provided multi-aperture 
K-photometry at 3.6, 5.3, 7.2 and 9.3\arcsec\  diameters. The origin of 
the K-band luminosity in nuclear starbursts is one of the key aspects
investigated by Devereux. 
Both the old and intermediate, and even the young populations,
can provide a significant contribution to the K-band luminosity. 
These three populations have observationally distinguishable
stellar features in the optical spectrum, especially 
when looking at the Balmer discontinuity and the prominence of hydrogen 
and metallic absorption lines. With our synthesis models we can thus check the
consistency of the  contribution of the different populations in the 
near infrared and in the optical, simultaneously.

In order to homogenize the data by Devereux with ours, we have calculated
the fraction of the K-band flux that falls within our slit aperture.
To this purpose we have computed the  surface brightness
profile that fits the multi-aperture data. The data are consistent with
exponential luminosity profiles, with scale lengths between 1--1.5\arcsec.
This suggests that the source of the K-band luminosity is the stellar disk
rather than a very compact source. The maximum luminosity allowed for a compact 
source is $\approx50$\% of that observed within the 3.6\arcsec\  aperture.
From the fitted surface brightness profile, we
calculated the K-band flux that falls within our slit aperture, 
and then computed the corresponding ratios with the visual flux of our 
spectra, F$_{\rm K}$/F$_{\rm V}$. The resulting values are tabulated in 
column~8 of Table \ref{table1}.

\begin{figure*}[htb]
\figurenum{1}
\epsscale{1.2}
\plotone{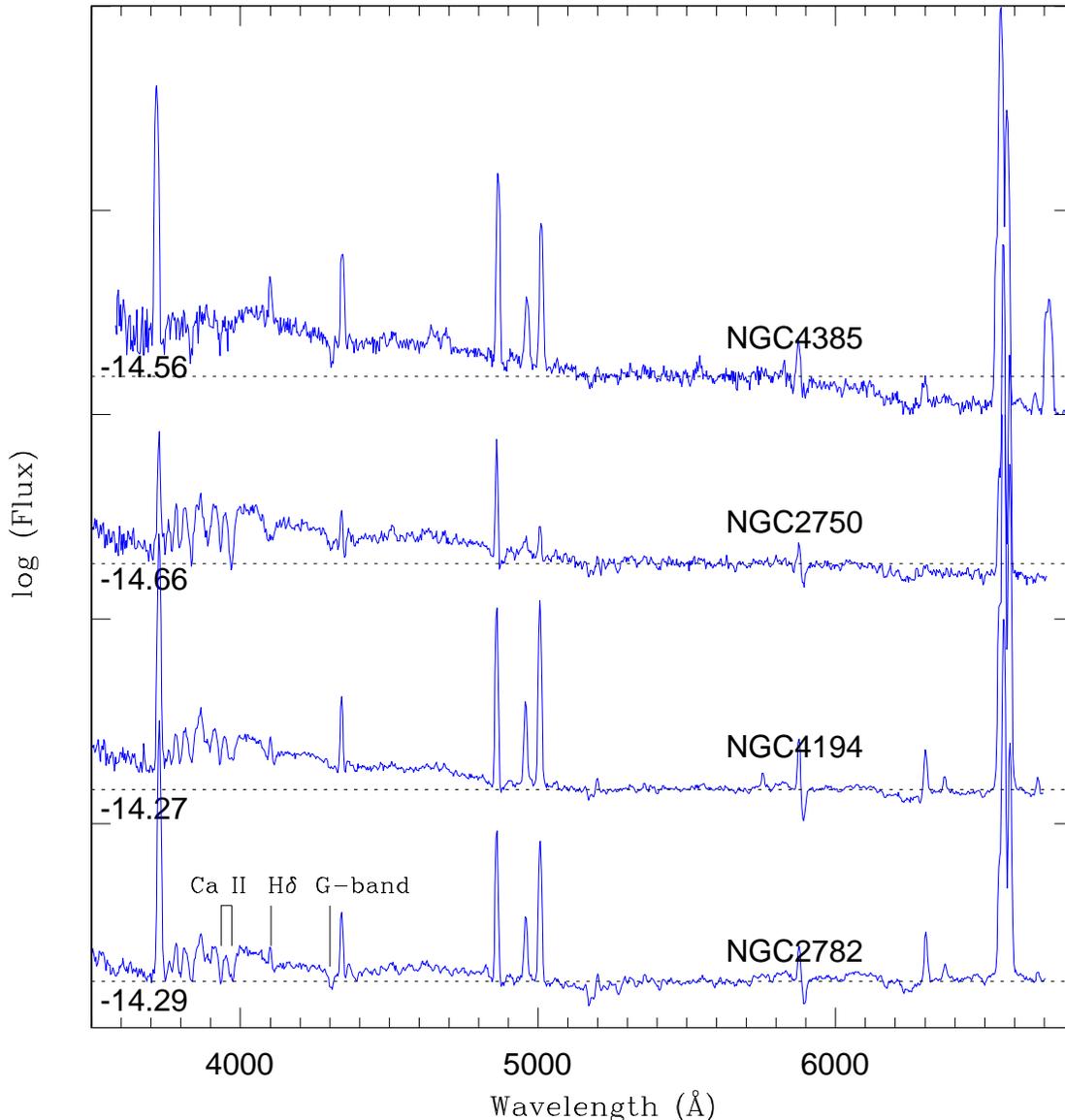}
\vspace*{-2cm}
\caption[]{Observed spectra of the starburst nuclei of our sample galaxies.
The spectra are arranged in three panels with optical colors becoming 
increasingly redder from NGC\,4385 to NGC\,4818. The flux scale for each plot
is shown by the dotted horizontal line. The value adjacent to this line
is the log of the visual flux (erg\,cm$^{-2}$\,s$^{-1}$\AA$^{-1}$) and
the vertical tick marks are placed at each 0.5 dex.
The position of H$\delta$ and the prominent metallic lines are marked.
The red part of the spectrum is dominated by the H$\alpha$ emission line,
while in the blue part several metallic absorption lines can be seen.
Absorption line strength gradually dominates over the emission line 
strength as we move towards higher order Balmer lines, with the H$\delta$ 
line seen only in absorption in the majority of the starbursts.
}
\label{fobs}
\end{figure*}

\begin{figure*}[htb]
\figurenum{1}
\epsscale{1.2}
\plotone{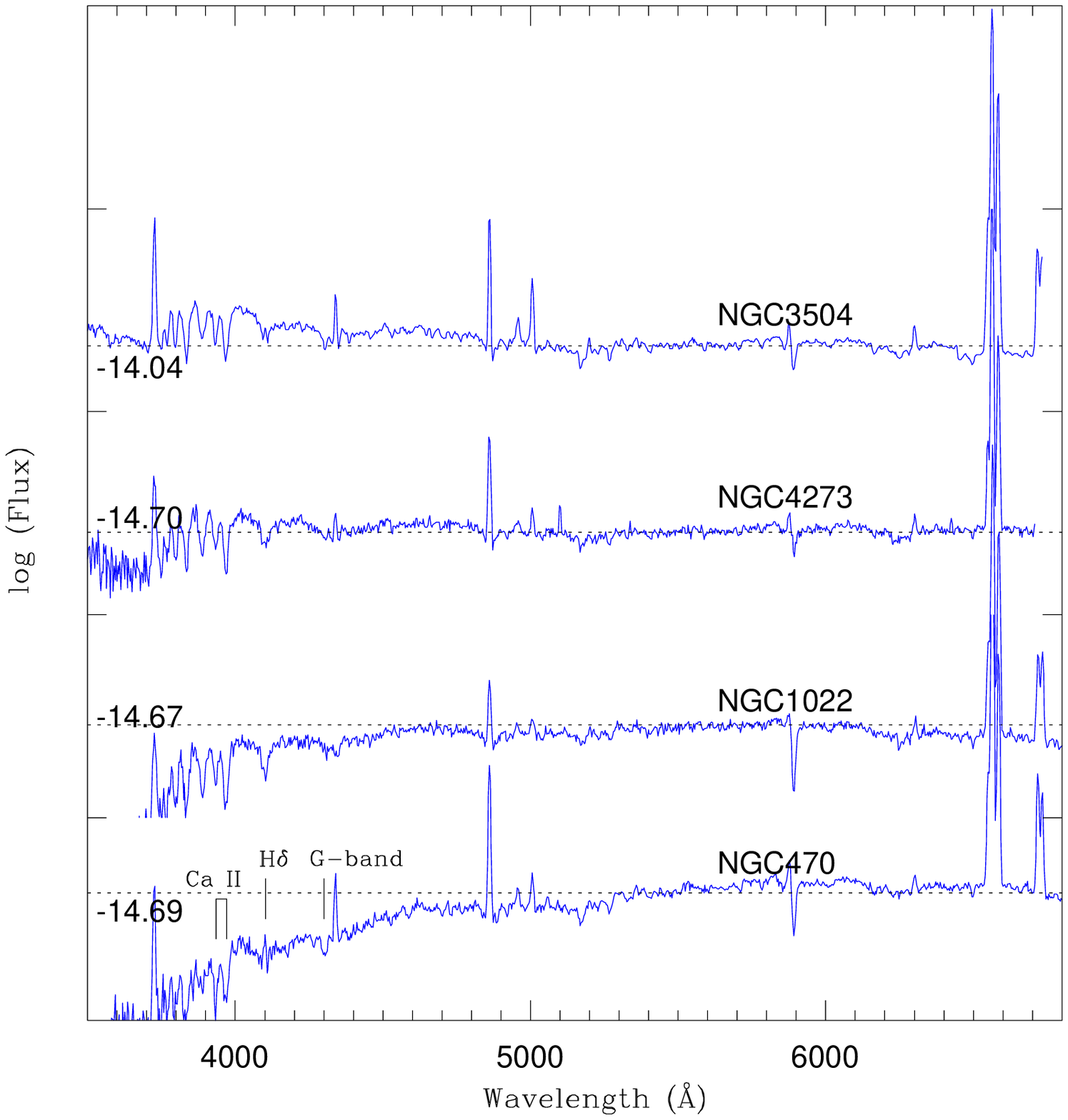}
\caption[]{Continued}
\end{figure*}

\begin{figure*}[htb]
\figurenum{1}
\epsscale{1.2}
\plotone{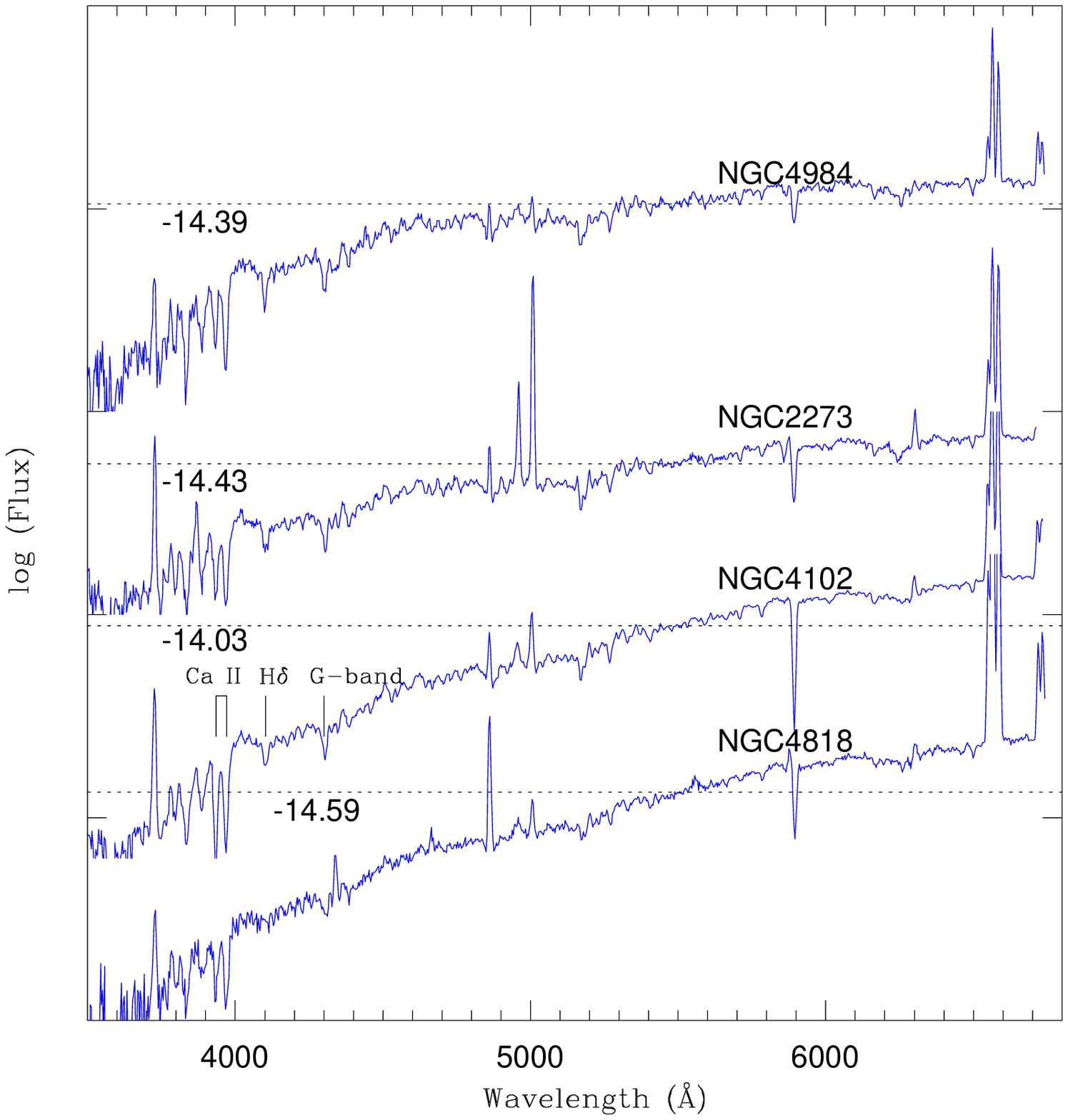}
\caption[]{Continued}
\end{figure*}
\section{Results and Discussions}

The observed spectra of the sample galaxies are shown in Figure~1.
These spectra indicate a wide range of optical properties
in spite of having nearly similar NIR and FIR luminosities.
Spectra are arranged according to their colors in order to easily visualize
the gradual change from blue (NGC\,4385) to red (NGC\,4818) starbursts. 
The most noticeable characteristic of these spectra is that the H$\alpha$ and 
H$\beta$ lines are seen in emission, while the higher order lines of 
the Balmer series are seen in absorption. 
This characteristic behaviour was also present in the sample of Very 
Luminous Infrared Galaxies analyzed by Poggianti et al. (2001).
The transition from emission to absorption happens 
at H$\delta$, with 11 of the 12 galaxies having the absorption feature  
stronger than the emission line. 
The only exception is the Wolf-Rayet galaxy NGC\,4385 (Mrk\,52). Besides, 
several metallic lines, notably the \caiih\ \& K lines are also seen in 
absorption, even in NGC\,4385. 
The simultaneous presence of absorption and emission lines is clearly an 
indication that more than one population contributes to the observed spectrum. 

To disentangle the contribution of the different stellar components, and
also to investigate the role of dust in reddening the spectrum,
especially in red starbursts, we have 
applied the model described in the previous section to our starburst spectra.
In order to highlight the success of the method, we summarize here the
salient characteristics of the different stellar populations.
The Balmer emission lines suggest the presence of a young population rich in 
OB stars. The intermediate age stellar populations are 
characterized by the prominence of hydrogen absorption lines and by a
strong gradient around 3700\,\AA. They do not show the discontinuity  at 
4000\,\AA\ (Bruzual 1983), which is instead a characteristic feature of 
the old populations.

Between 3700\,\AA\  and 4000\,\AA\  the spectral shape of the intermediate age 
stellar populations is thus very different from that of the old populations. 
At wavelengths longer than 4000\,\AA, the  continuum of the intermediate 
age populations is characterized by the absence of typical absorption bands 
(e.g. Mg and G-band)
shown by the old populations (age $>$ 1 Gyr). Thus, a significant 
contribution of the old populations not only changes the equivalent width
of the hydrogen lines (emission or absorption) but also modifies the shape 
of the continuum between 3700\,\AA\  and 4000\,\AA\  and introduces visible 
features in the optical region.

Another noticeable characteristic of the old populations is the depth of 
the \caiik\  line at $\lambda\simeq$ 3934\,\AA. This line is almost 
absent in young and intermediate age populations. It becomes prominent 
at ages greater than a few Gyrs and thus it is a strong indicator of the 
dominance of the old populations. On the contrary the \caiih\  line is 
blended with the H$\epsilon$ line and may appear strong even for intermediate
age populations. In the present modeling, we have not used the \caiik\  
line as a constraint, because of lack of information on its behaviour 
at different metallicities. Indeed the atmosphere models used to 
compute SSPs are based on solar metallicity stars and only the 
indirect effect of metallicity on the turnoff of the isochrone can be modeled.
Recent advances in atmosphere models will allow the
inclusion of this constraint in the near future (Rodriguez-Merino et al. 2003).
However, though not included as a constraint, we may notice that the \caiik\ 
and \caiih+H$\epsilon$ absorption lines are fairly well reproduced by the 
models, providing a further strong support to the validity of the solution.

In summary, when a good fit to the data is reached, the solution is 
unique as far as the three broad age ranges are considered. More specifically, 
while almost equally good solutions can be obtained by 
interchanging the different populations within a given episode (young, 
intermediate or old), it is not possible to find a consistent
solution by suppressing or adding an entire episode of star formation.

\begin{figure*}
\figurenum{2}
\centerline{\psfig{file=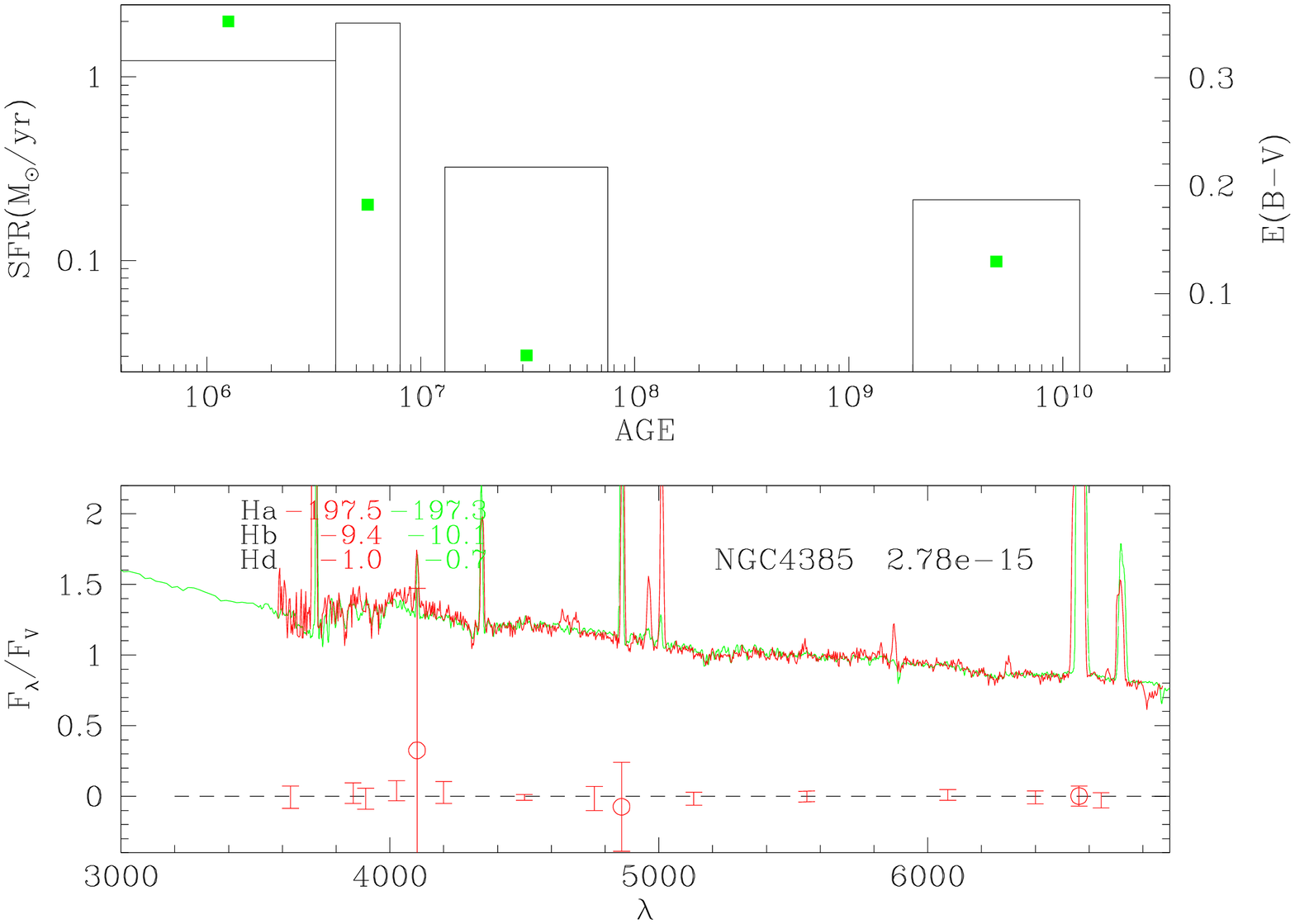,width=16.0cm,angle=-90}}
\vspace{-1cm}
\caption[]{Comparison of the observed nuclear spectra of 3 representative 
program galaxies with the model spectra (bottom panel). 
The residuals (observed$-$model) at each of the fitted bands are shown below 
the spectrum along with the error bars, with the circled error bars 
corresponding to the residuals at the line bands. 
The inferred star formation history is shown by the histograms in the upper 
panel. The filled squares correspond to the color excess E$_{\rm B-V}$
(numbers on the right-axis) for each star formation episode.
Fits for NGC\,4385, NGC\,1022 and NGC\,4984 are shown. These three 
galaxies represent blue, intermediate color and red starbursts respectively.
}
\label{f2}
\end{figure*}

\begin{figure*}
\figurenum{2}
\vspace{-1.5cm}
\centerline{\psfig{file=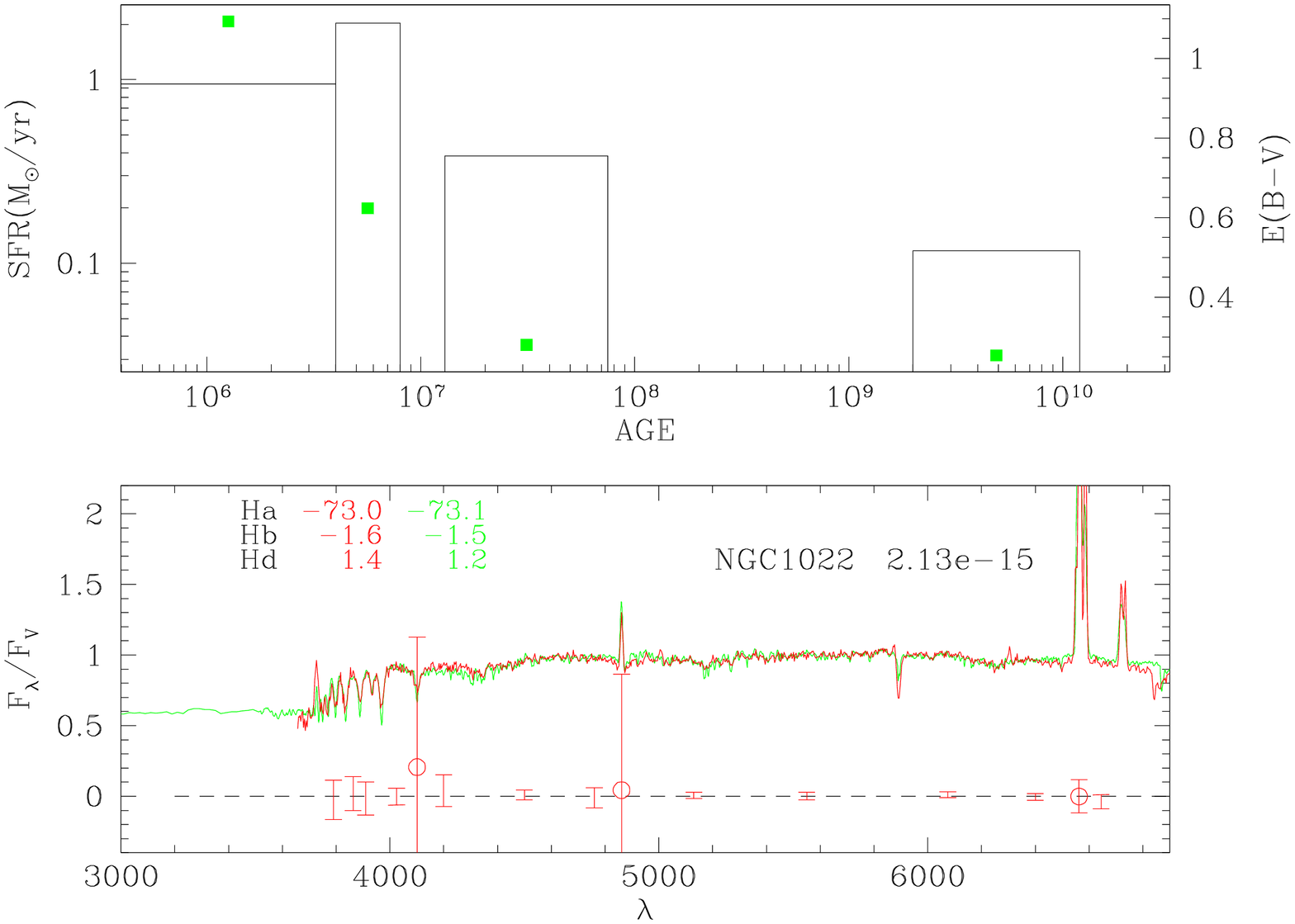,width=16.0cm,angle=-90}}
\centerline{\psfig{file=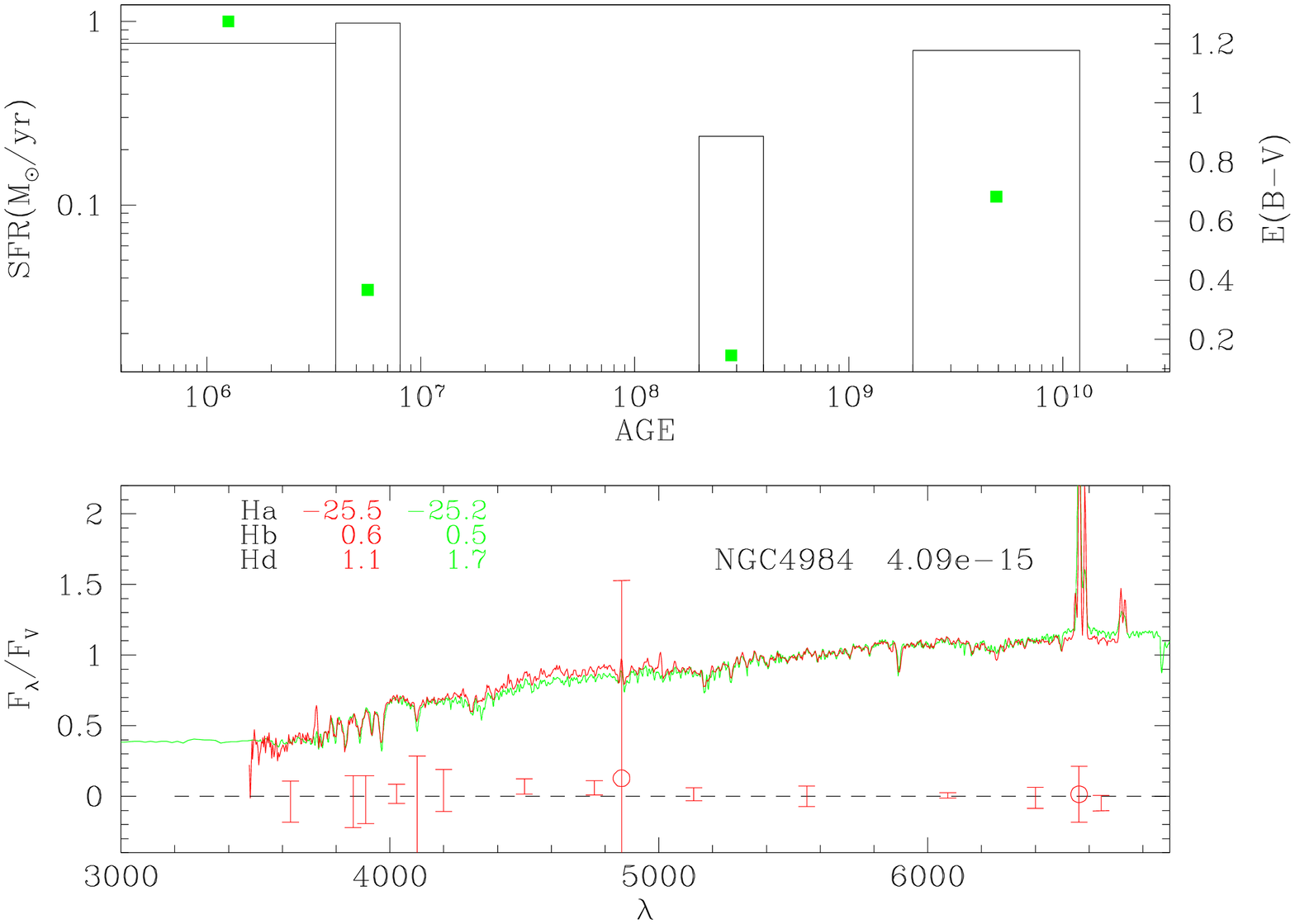,width=16.0cm,angle=-90}}
\caption[]{Continued} 
\end{figure*}
\begin{center}
\begin{deluxetable*}{rcccccccccrcr}
\tabletypesize{\scriptsize}
\tablewidth{0pc} 
\tablecaption{Star Formation and reddening properties of the starburst nuclei}
\tablehead{
\colhead{NGC} 
              & \multicolumn{4}{c}{Young Burst}  
              & \multicolumn{3}{c}{Intermediate}  
              & \multicolumn{2}{c}{Old} 
              & \colhead{FIR/V} 
              &  \colhead{F$_{\rm K}$/F$_{\rm V}$}  & \colhead{Mass} \\
\colhead{   } 
              & \multicolumn{2}{c}{(0--4 Myr)}  
              & \multicolumn{2}{c}{(4--8 Myr)}  
              & \multicolumn{2}{c}{}  
              & \multicolumn{1}{c}{Age}  
              & \multicolumn{2}{c}{(Bulge or Disk)} 
              & \colhead{} & \colhead{} & \colhead{$\times10^9$} \\
\colhead{} 
 & \colhead{SFR} & \colhead{E$_{B-V}$} 
 & \colhead{SFR} & \colhead{E$_{B-V}$} 
 & \colhead{SFR} & \colhead{E$_{B-V}$} & \colhead{Myr}
 & \colhead{SFR} & \colhead{E$_{B-V}$} 
 & \colhead{Model} & \colhead{Model} & \colhead{\msun} 
}
\startdata
470 & 1.78&0.73&4.46&0.66& 1.91& 0.49&300&0.39& 0.40&21.88& 0.37 &4.35\\
1022& 0.95&1.09&2.04&0.62& 0.39& 0.28&50 &0.12& 0.25&27.83& 0.33 &1.21\\
2273& 3.77&1.28&0.15&0.00& 0.34& 0.00&50 &1.62& 0.55&22.86& 0.45 &16.26\\
2750& 2.95&0.98&1.90&0.00& 0.35& 0.00&50 &0.52& 0.33&17.00& 0.23 &5.25\\
2782& 4.26&0.65&3.48&0.00& 0.52& 0.00&50 &2.07& 0.51&12.49& 0.32 &20.73\\
3504& 8.24&1.14&2.62&0.00& 0.30& 0.00&100&1.06& 0.44&34.08& 0.29 &10.65\\
4102& 7.25&1.78&0.00&2.53& 0.10& 0.00&50 &1.01& 0.66&100.18&0.70 &10.10\\
4194&18.32&1.05&5.49&0.00&0.45& 0.00&300&1.51& 0.53&47.42&0.29  &15.34\\
4273& 0.58&0.92&0.19&0.00&0.09& 0.00&50 &0.18& 0.41&15.90& 0.28 & 1.79\\
4385& 1.22&0.35&1.95&0.18&0.32& 0.04&50 &0.21& 0.13& 9.86& 0.16 & 2.16\\
4818& 5.08&1.59&0.08&0.00&0.02& 0.00&300&0.43& 0.68&154.38&0.72 & 4.34\\
4984& 0.77&1.27&0.99&0.37&0.24& 0.15&300&0.70& 0.68&18.25& 0.57 & 7.09\\
\hline
\enddata
\label{table2}
\end{deluxetable*}
\end{center}
For purpose of conciseness, we show in Figure~\ref{f2} only three model
fits, while the results for all the galaxies are presented in Table~2.
NGC\,4385 and NGC\,4984 are selected as representative blue and red 
starbursts, respectively. NGC\,1022 represents an intermediate case.
For each galaxy, the bottom panel of the figure shows the best fit model
(lines extending shortward of 3500\,\AA) superimposed on the observed spectrum. 
The spectra are normalized to 
their $V$-band luminosity. The residuals (observed $-$ model) and the
estimated errors at the 14 fitted points are shown below the corresponding 
features. The residuals on the H$\alpha$, H$\beta$ and H$\delta$ line bands
are identified by circled error bars.
The histogram in the top panel represents the star formation 
history (in \msun\,yr$^{-1}$) with the relevant SFRs on the left axis. 
The filled square at each histogram position represents the color 
excess \ebv\  for each population. The \ebv\ scale is shown on the right axis.

Models reproduce the observed spectra fairly well, as can be judged by
the low residuals at the fitted continuum bands. The fits were even able to 
reproduce the strength of some of the spectral features, such as the 
\caiih\ and K absorption lines, that were not included in the merit function.
The relatively large 
residuals and errors on H$\beta$ and H$\delta$ lines are due to their
small equivalent widths caused by the stellar absorption lines almost 
wiping out the nebular emission lines. Observed (left column) and 
model EWs of H$\alpha$, H$\beta$ and H$\delta$ are written on the plots.
We did not try to fit the emission lines of [\ion{O}{2}] and [\ion{O}{3}], 
because their intensities depend strongly on the physical conditions 
in addition to the strength of the stellar populations.

For each starburst, the model has found four populations
--- two young populations, one intermediate and one old population, 
each with its distinct mass (or SFR) and extinction. 
Two young populations may represent either two short-duration bursts, 
one older than the other, a continuous star formation lasting for 
more than 4~Myr,  
or two bursts of nearly the same age, but with vastly differing extinction.

\begin{figure*}
\figurenum{3}
\epsscale{1.3}
\plotone{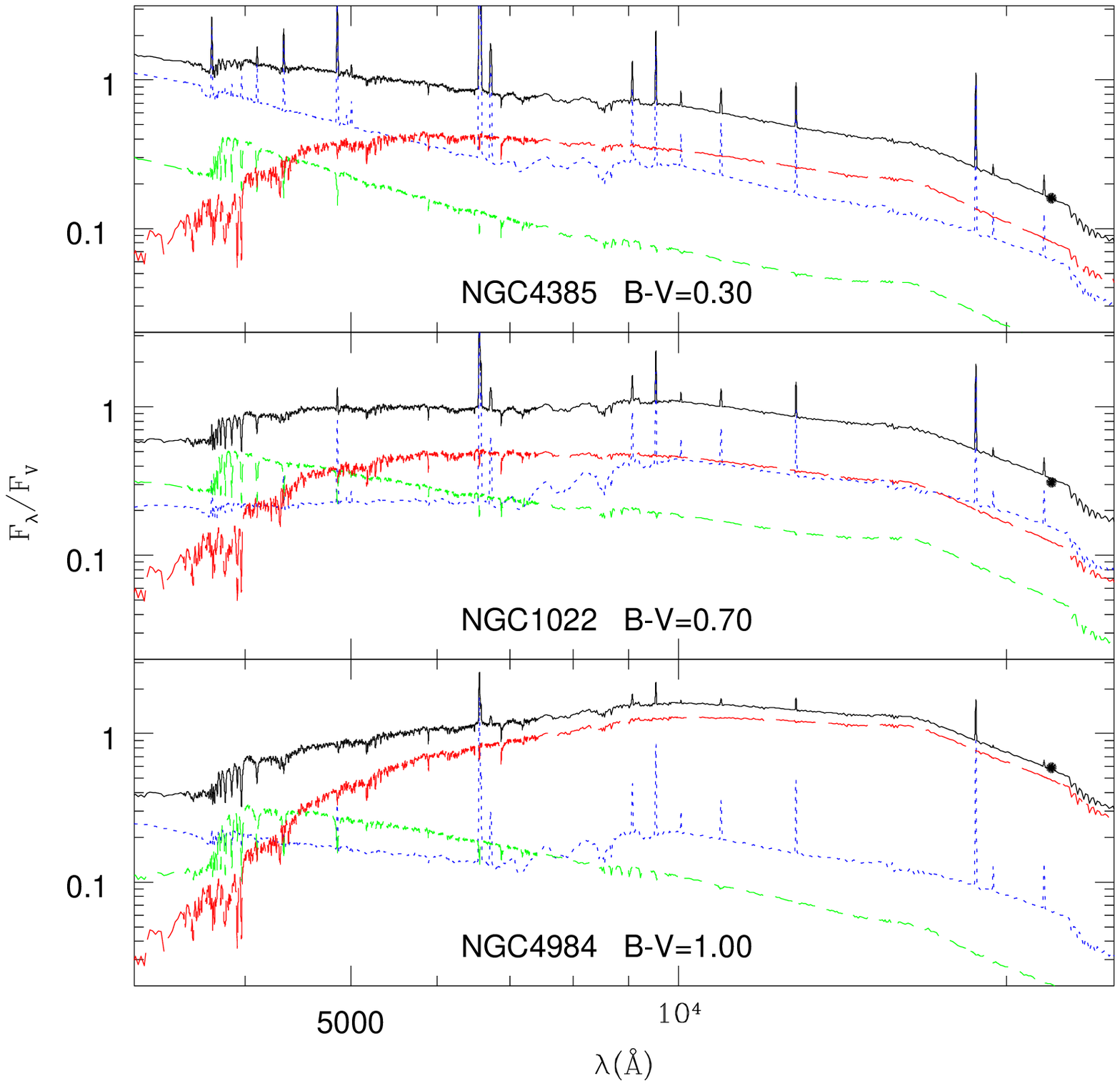}
\vspace{-1.5cm}
\caption[]{Decomposition of the best-fit model spectrum (black, solid line)
into spectra of young (dotted), intermediate (short-dashed) 
and old (long-dashed) populations, for 3 representative sample galaxies. 
NGC\,4385 is the bluest of the sample galaxies, while NGC\,4984 is among
the reddest of the sample galaxies. The thick dot at the
2.2$\mu$m  is the estimated K-band flux (normalized to the V flux) 
within the spectroscopic slit. 
The blue part of the spectrum is dominated by the young and intermediate 
populations, while the red and NIR parts are dominated by the old population.
}
\label{mod}
\end{figure*}
The results of the fit for all the galaxies are presented in Table~2.
Columns 2--10 contain the model derived SFR and the reddening 
\ebv\ for the four selected populations. For the intermediate-age
population, we also tabulate the age of the best-fitted SSP in Column~8.
Model derived ratios, FIR/V and F$_{\rm K}$/F$_{\rm V}$ are given in 
columns 11 and 12 respectively.  The last column contains the total 
mass sampled by the slit with the adopted IMF.

A comparison of the computed FIR/V values and the observed values
(Columns 6 and 7 of Table~1), indicates that the two values are in the same
range, taking into account the differences in
the optical and FIR apertures, discussed in Section~3.2.
It may be recalled that in Very Luminous Infrared Galaxies, which on 
average are 10 times more luminous than our galaxies, observed values
are systematically higher than the model values, suggesting a significant
amount of star formation hidden from the optical view (Poggianti et al. 2001).
The fact that the observed values are comparable with the predictions
of the model in our sample of galaxies, implies that most of the FIR emission
detected by IRAS originates in the starburst nucleus and that
the starburst regions do not contain a significant amount of
optically hidden star-forming regions.

\begin{figure*}
\figurenum{4}
\vspace{-2cm}
\centerline{\psfig{file=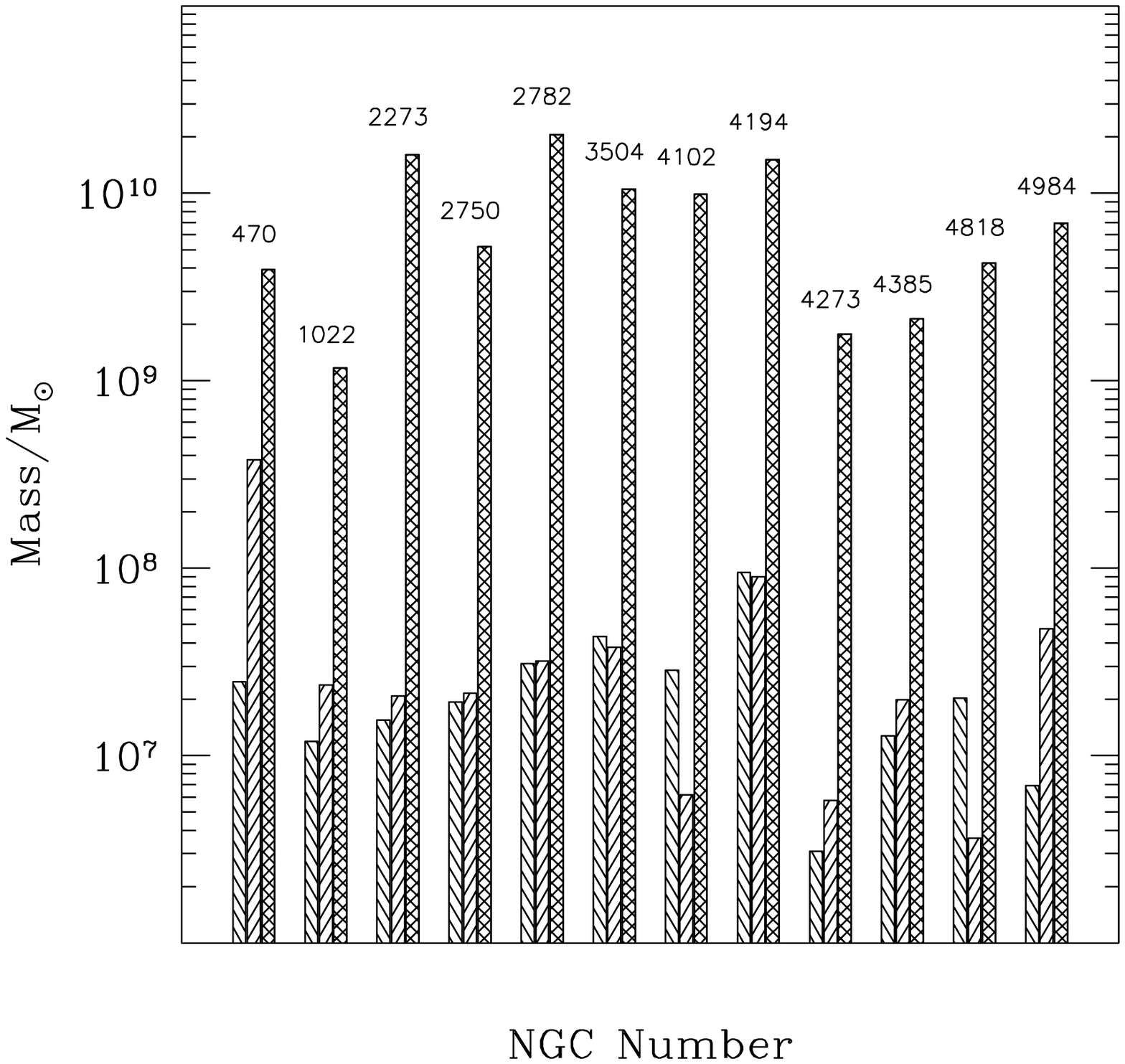,width=16.0cm,angle=0}}
\caption[]{Masses of the young  (left vertical bar),
intermediate age (center vertical bar) and old (right vertical bar) component,
sampled by our slit. While the first two masses are within a factor of two
of each other in most cases, the old population mass is up by as much 
as a factor of 100.
}
\label{hmas}
\end{figure*}
An insight into the relative importance of the young, intermediate-age and old
populations in the overall history of nuclear star-formation
can be obtained by analyzing the spectra of individual populations. 
In Figure~\ref{mod}, we plot the best-fit model spectrum along with the 
spectra of individual populations for the three representative galaxies.
Observed K-band photometry is also displayed in these
plots. Note that the plotted spectra are attenuated by
the extinction values obtained for each population.

In the following sub-sections, we discuss in detail the mass and 
extinction of the different populations.

\subsection{Masses of the distinct populations}

In the presence of selective extinction, deriving masses of different 
populations is not straight forward. While the intrinsic mass-to-light 
ratio is the 
basic ingredient of any population synthesis technique, the option to vary
attenuation for each population, introduces a new degree of freedom. 
Unreasonable high masses can be obtained by letting the attenuation to 
increase with the strength of an intermediate or old stellar population. 
The algorithm described in Section \ref{models} automatically copes
with this degeneracy by selecting the minimum mass-to-light ratio
solution.

\begin{figure*}
\figurenum{5}
\centerline{\psfig{file=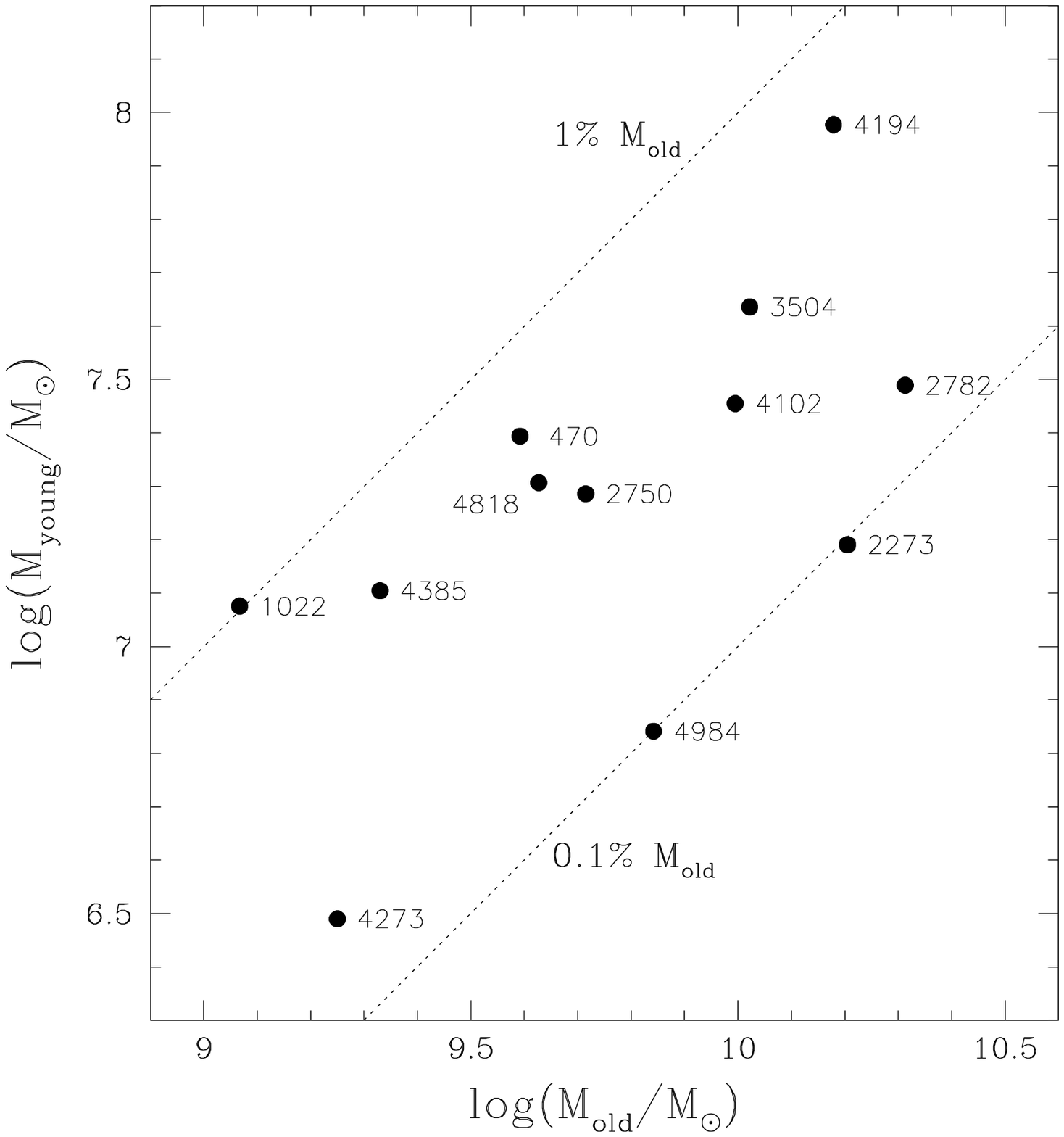,width=16.0cm,angle=0}}
\caption[]{Mass of the young component (proportional to the
current SFR) vs. mass of the old component. Dotted lines correspond to the
boundaries where the young burst mass is 0.1 and 1\% of the total mass.
The observed trend implies that the young mass is proportional to the old mass.
 }
\label{momy}
\end{figure*}

Mass, rather than SFR, is the fundamental output of the model.
The SFR of each population, tabulated in Table~2, is obtained by dividing
the mass by the assumed duration of star formation (see Section 3.1).
The derived mass of each population is 
plotted as a bar histogram in the upper panel of Figure~\ref{hmas}.
For each galaxy, the histograms from left to right denote the masses
of young, intermediate-age and old populations. It can be seen that
the mass of the young population is comparable to that of the
intermediate-age population in most cases, whereas the mass of the
old population is significantly higher.
The mass of the young population is plotted against the old stellar mass
in Figure~\ref{momy}. The dotted lines in this plot delimit the zone 
where the young stellar mass is 0.1\% and 1\% of the old stellar mass.
All the points lie between these two limits. This indicates that the 
recent starburst contributes a tiny amount to the overall visible mass
of the nucleus. It is interesting to note that corresponding fraction 
is 5\% in blue compact galaxies (P\'erez-Gonz\'alez et al. 2003). 
The inferred total stellar masses of the starburst regions are in 
the range of masses that are usually derived for the galactic bulges.

\subsubsection{Present vs Past SF in starburst nuclei}

The ratio between the current star formation rate (in the last 8~Myr) 
and the average past star formation rate, SFR/$<$SFR$>$, often referred to 
as the birthrate parameter (Kennicutt, Tamblyn \& Congdon 1994), is a 
useful parameter to measure the strength of the most recent burst.
The value of this ratio is less than 0.1 for normal early-type disk galaxies, 
which are the Hubble types that dominate our sample.
Our sample of starbursts have birthrate parameter in the range 1--12,
with an average value of $\simeq$5.0$\pm$3.3.
This reiterates that the galaxies of our sample are indeed starburst galaxies.

All starburst regions are found to have an intermediate-age stellar
population. Unfortunately, it is difficult to distinguish differences in
spectral features of an evolving starburst between $\sim$15 and 500~Myr. 
An evolved spectra of an intense short-duration burst will be 
indistinguishable from a lower intensity continuous star formation process. 
There can be three possible origins for the intermediate-age population:
(1) it represents stars formed earlier in the current star formation episode, 
(2) it represents the exponential tail of the disk star formation history,
(3) it represents an episode of star formation similar to the present one, 
but that took place around a few hundred million years ago.
In the following paragraphs we discuss these possibilities in more detail.

In majority of our sample galaxies, the best-fit SSP that represents the 
intermediate age population has an age of 50~Myr (see Column 8 in Table~2). 
Considering that the SED almost remains unchanged between 15--75~Myr,
the intermediate age population could be as young as 15~Myr.
Case (1) would then
imply that the star formation is going on un-interrupted for at least 15~Myr.
The histograms of Figure~2, suggests that there is
a gap in star formation between 8 and 15~Myr. This gap corresponds to the
populations of red supergiants, which are also represented by the 7~Myr SSP.
Hence equally good solutions could have been obtained by extending the duration 
of the young burst upto 15~Myr. The average present SFRs, in that case, 
would be a factor of two lower than those presented in Table~2.
If the SFR remains constant, then the mass of the processed stars
scales proportionately with the duration of the burst.
We have seen in the previous section
that the mass of the intermediate age population is comparable to that of 
the young population. This implies that the duration of the intermediate
age population is comparable to that of the young population ($\sim15$~Myr).
The duration would be even smaller if the SFRs were higher at earlier times.
Hence the maximum duration of the present episode for a continuous star 
formation model is $\sim$30~Myr.

If case (2) above is correct, then the masses of the populations would
be proportional to the life-time of the stars constituting the populations.
This would imply that the mass of the intermediate-age population is
around 1\% of that of the old population. 
Star formation rate in the disks of early-type galaxies decreases exponentially 
with time and the expected mass would be even less. The derived masses of
the intermediate-age populations are indeed $\le1$\% of that of the
old population, and hence we cannot rule out the possibility (2).
The third possibility, case (3), is also plausible because
the mass of the intermediate-age population is comparable with that of
the present burst. 

Perhaps a way to distinguish the case (2) from the other two cases is to
look at the spatial distributions of the stellar populations.
In case (2), stellar populations are expected to be uniformly
distributed whereas in cases (1) and (3) old star clusters are expected.
The fact that SSCs are frequently encountered in starburst regions and
that some of them are of intermediate age and spatially separated from the
present active knot, makes us believe that case (3) is the most likely 
possibility. It is also possible that all the three processes of formation of
intermediate-age population are at work in starburst regions. The relative
importance of the three processes might vary from one galaxy to another.

\subsection{The Origin of K-band luminosity}

The origin of the K-band luminosity in nuclear starbursts has been 
the subject of 
debate ever since the first starburst sample was defined by Balzano (1983). 
She found that the near infrared colors as well as the K-band 
luminosities, were unaffected by the starburst activity. 
This can be seen in Table~3, where we have tabulated Balzano's H$-$K
and J$-$K colors of starburst and bright NGC galaxies in columns 2 and 3
respectively. NIR colors of starburst nuclei are indistinguishable
from those of the nuclei of normal galaxies. Cizdziel et al. (1985)
supported Balzano's findings, by analyzing the K-band growth 
curves of starburst nuclei, which indicated a lack of central concentration. 
On the other-hand, Rieke et al. (1980, 1993) interpreted the entire K-band 
luminosity in the archetype starburst M\,82, as due to the red supergiants 
associated with the present starburst activity. 

\begin{figure*}
\figurenum{6}
\centerline{\psfig{file=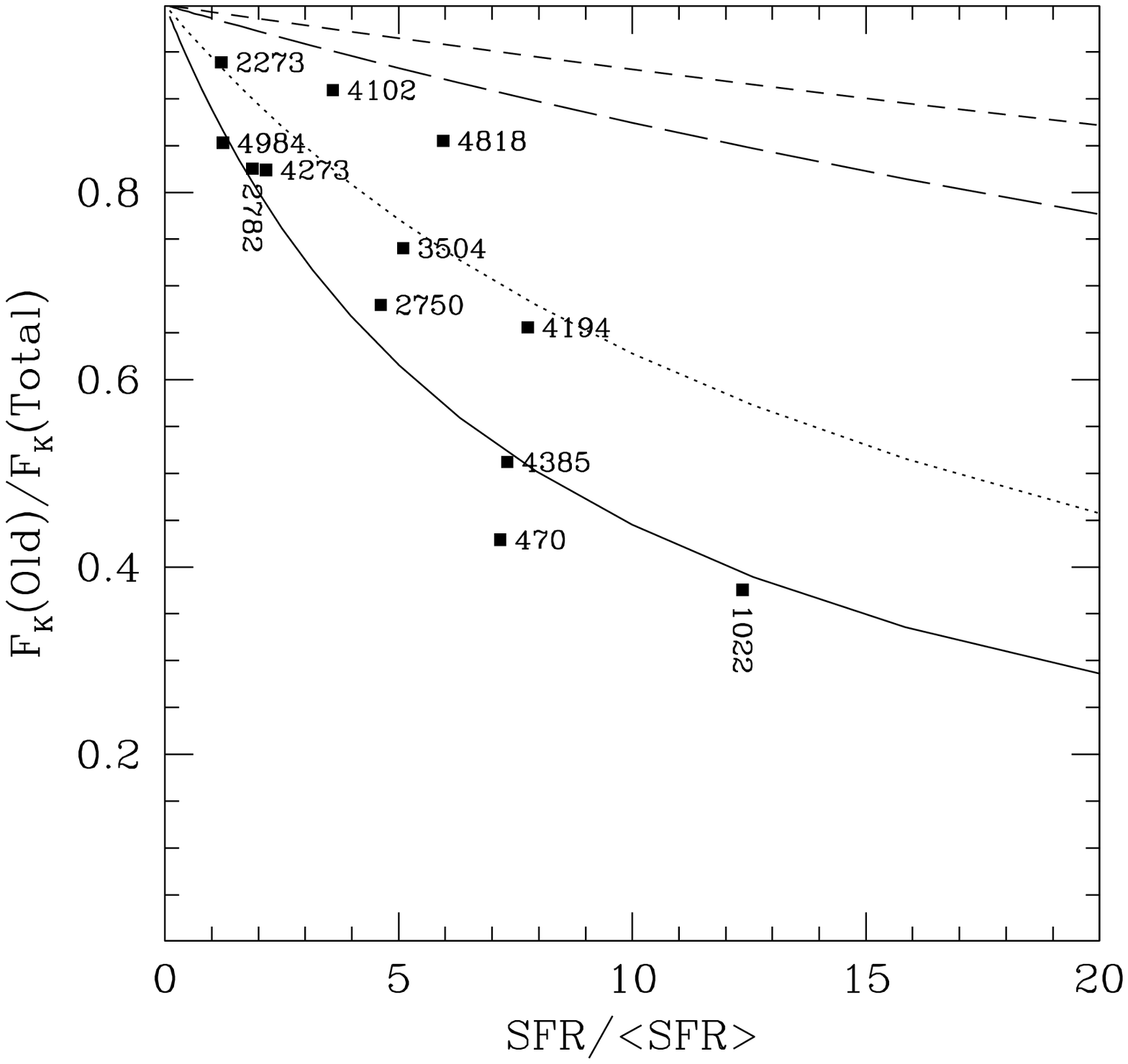,width=16.0cm,angle=0}}
\caption[]{Fractional contribution of the old stars to the K-band luminosity
plotted against the birthrate parameter. As expected, the contribution
of the old stars to the K-band decreases with increasing birthrate. 
In three quarters of the sample galaxies, the fraction remains above 60\%.
Lines represent theoretical predictions using SSPs and refer to
a) solid line: Z=2.5\zsun, young age 10\,Myr and old age 10\,Gyr;
b) dotted line: Z=\zsun/5, young age 10\,Myr and old age 10\,Gyr;
c) short dashed line: Z=2.5\zsun, young  5\,Myr and old  10\,Gyr;
d) long dashed line: Z=\zsun/5, young  5\,Myr and old  10\,Gyr.
 }
\label{lksb}
\end{figure*}

Our sample of starburst galaxies is well-suited to resolve
the above controversy. Devereux (1989), from where our
sample was chosen, discussed the existence of a correlation 
between the K-band and the 10$\mu$m luminosities, for galaxies that are 
luminous at 10$\mu$m. Since 10$\mu$m emission is related to the recent 
star formation, this correlation implies that the K-band emission is also 
related to the recent star formation.
On the contrary, by means of our detailed population synthesis technique,
we find that, in the same galaxies showing the correlation between 
10$\mu$m and K-band luminosity, a major fraction of the K-band emission 
comes from the old population. A strong contribution to the K-band 
luminosity from the young and/or intermediate age stellar populations 
is excluded by the overall fit of the broad continuum features and 
inspection of some metallic lines (e.g. \caiik) of the optical spectrum.
To better clarify this point, 
we plot in Figure~\ref{lksb}, the fractional contribution of the old stars
to the total K-band luminosity against the birthrate parameter. 
We also show a comparison with theoretical predictions based on
simple stellar population, assuming that the old stars have 
an average age of 10\,Gyr. Calculations were done for two ages of the
young component: 10\,Myr and 5\,Myr, which represent phases of starburst
with and without the red supergiants respectively. The metallicities for the
young and old components were assumed to be the same.
The age of the young component and the metallicity for the different 
plotted lines are:
a) 10\,Myr and Z=2.5\zsun\ (solid line);
b) 10\,Myr and Z=\zsun/5 (dotted line);
c) 5\,Myr and Z=2.5\zsun\ (short dashed line);
d) 5\,Myr and Z=\zsun/5 (long dashed line), respectively.

As expected, the contribution of the old population decreases with 
increasing birthrate, but only for a birthrate parameter larger than 
about 7 this contribution falls below 50\% of the total light.
We have only two cases where this contribution is less than 50\%,
NGC\,470 (42\%) and NGC\,1022 (38\%). NGC\,1022 is among the galaxies 
with the lowest SFRs in our sample (1\,\msun\,yr$^{-1}$), 
and its high birthrate is due to the low mass of the old population. 
This is to stress once more that
the relative contributions of the young and the old populations
to the K-band depends on the birthrate parameter rather than the present SFR.
The figure also illustrates the dependence of the contribution of the old 
population, on the age of the dominant young population and the metallicity. 
For a given birthrate, the contribution of the old population is higher
for lower age and lower metallicity of the young population. It can be
inferred from the theoretical curves in the plot that for young bursts
which are yet to evolve into the red supergiant phase, more than 80\%
of the K-band flux originates from old stars, unless the birthrate
parameter is much higher than the plotted range.

Notice that a lower contribution from the old population
in NGC\,1022 may also be inferred from the relatively 
low strength of the \caiik\ line (Fig.~1).

Further insight about the nature of the K-band emission may be obtained by 
analysing the radial intensity profiles. As discussed in Sec.~3.3,
the Devereux multi-aperture K-photometry, is consistent with exponential 
luminosity profiles (scale-lengths between 1--1.5\arcsec), with the compact 
nucleus not contributing more than 50\% within the 3.6\arcsec\  aperture.
The contribution of the compact nucleus would be even less if the real 
intensity profiles of program galaxies are $r^{1/4}$ bulge profiles rather 
than the exponential ones that we have assumed. This disagrees with Devereux 
claim of ``central concentration'' of K-band flux, in these galaxies. 
We however find an evidence of systematic reddening of V$-$K color towards 
the center, which is expected in all disk galaxies due to the flattening of 
the disk at shorter wavelengths due to optical depth effects in {\it dusty} 
disks (Burstein, Haynes \& Faber 1991; Evans 1994).

In the light of this new analysis, we may ask what is the origin of the 
correlation found between the K-band luminosity and the 10$\mu$m luminosity, 
in these galaxies. Figure~\ref{momy} helps us to clarify this point.
In this figure, the mass of the young component, which is directly 
proportional to the star formation rate, is compared with the mass 
of the old component, essentially the total mass of the bulge. 
There is a trend for the more massive galaxies to host a stronger nuclear 
starburst.

Thus, the relation between the K-band and the 10$\mu$m luminosity can be 
essentially ascribed to the underlying relation between the total galaxy 
mass (hence the mass of the old component and the K-band luminosity) and 
the gas mass (hence star formation rate and  10$\mu$m luminosity). 

This explanation is also supported by the fact that Devereux's M\,82-like 
starburst sample galaxies are earlier than Sb (and hence relatively bigger), 
while the mean Hubble type of Balzano's sample is Sc. On the other hand 
the observed lack of high 10$\mu$m emission in faint K-band sources is 
basically due to the lower gas mass (not fraction) of low mass galaxies.
Thus the M\,82-like starburst sample of Devereux contains early type 
star-forming galaxies.

\begin{center}
\begin{deluxetable}{ccccc}
\tablewidth{0pc} 
\tablecaption{NIR colors of nuclear starbursts and normal galaxies}
\tablehead{
\colhead{Color} & \colhead{Balzano} & \colhead{Bright NGC} 
                & \colhead{M82-like} & \colhead{Young Burst} \\
}
\startdata
$H-K$ & 0.32 $\pm$0.12 & 0.31$\pm0.10$ & 0.34 $\pm$0.05 &  0.51 $\pm$0.17 \\
$J-K$ & 1.02 $\pm$0.12 & 1.04$\pm0.10$ & 1.08 $\pm$0.08 &  1.05 $\pm$0.14 \\
\hline
\enddata
\label{table4}
\end{deluxetable}
\end{center}
Interestingly, the $J-K$ and $H-K$ colors of the models that best fit the
observed spectra agree excellently with those reported for Balzano's 
sample of starburst galaxies (Balzano 1983). 
This is demonstrated in Table~3. Model colors of the total and only
the young burst are given in column~4 and 5 respectively.
It can be seen that the observed $H-K$ colors of starburst galaxies 
are indistinguishable from those of non-star-forming NGC galaxies. 

\begin{figure*}
\figurenum{7}
\centerline{\psfig{file=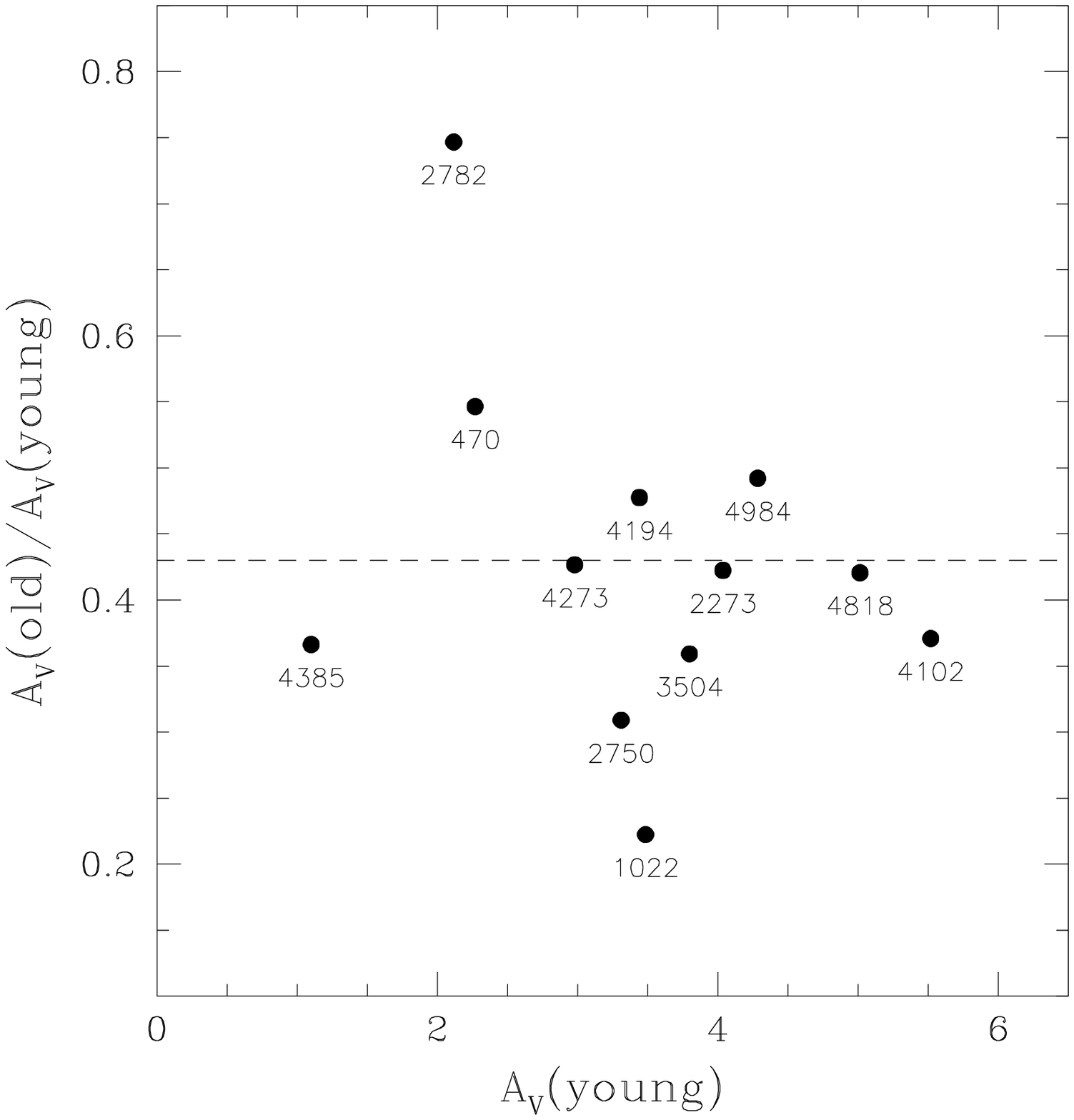,width=16.0cm,angle=0}}
\caption[]{Ratio of the visual extinction of the old to the young 
population, plotted against the visual extinction of the young population.
The dotted line is the mean value of the ratio. 
Note that the mean ratio is essentially the same as that obtained by
Calzetti et al. (1994). 
}
\label{ebv}
\end{figure*}
Thus most of the K-band luminosity of 
our starburst nuclei, seems to originate from old disk/bulge stars. 
However, we expect that in starbursts with a birthrate parameter 
that exceeds the average value found in our sample (about 5), 
the young stars begin to dominate the NIR fluxes. This may be the case
of high redshift starbursts with similar SFRs but with a significantly
lower amount of old stars.

\subsection{Extinction Pattern of Different Stellar Populations}

One of the aspects where our approach to spectral decomposition differs 
from the rest (e.g. Lancon et al. 2001) is in the treatment of extinction. 
In our starburst models, the extinction of each population is a free parameter, 
whose value is chosen to be consistent with both the continuum shape 
and the Balmer line ratios. 
In addition, the derived extinction does not depend on the extent of 
contamination by the underlying absorption, as the old populations
are simultaneously determined in the models.

The dots above the histograms in Figure~2 indicate the reddening value of 
each population. The reddening values, \ebv, of each population are also given 
in Table~2. It can be seen that the reddening of the young populations 
(age $\le$8~Myr) is always higher than that of intermediate-age and old
populations. Poggianti et al. (2001) also found a similar result for 
a sample of Very Luminous InfraRed Galaxies (VLIRGs). The reason for the 
higher extinction of the young population is well-known --- 
the young population is still embedded in its parent molecular 
clouds, which are known to be dusty (Silva et al. 1998).
Among the two young populations, the extinction of the youngest population
is systematically higher. This may suggest that stars
are already escaping from their parent molecular clouds at around 4~Myr.

The ratio of the extinctions of the old to the young population 
is plotted against the extinction of the young population in Figure~\ref{ebv}.
The latter value is calculated from the summed model spectra of the two
youngest populations (see description of Table~1 in Section 2 for more details).
The dotted line in the figure represents the average value of the ratio
for the sample galaxies, which is equal to $0.43\pm0.13$. 
Note that the rms of the ratio is only 0.13~mag, in spite of a range of more
than 4~mag for the extinction of the young population.
Thus the observed extinction towards the old stars is proportional to the 
extinction towards the young stars. Such a correlation is not expected 
if all the dust that causes the extinction of the old stars is 
mixed with the stars. Moreover the \ebv\ of the old population
in all cases, except NGC\,4385, is higher than 0.18~mag, the limiting
reddening for a uniform mixture of dust and stars (Poggianti et al. 2001).
All these arguments suggest that the same clouds that cause 
the extinction of the young population, are the principal contributors
to the extinction of the old stars. This could be understood
under a geometry in which the young star-forming regions are surrounded
by molecular clouds, and the majority of the old (bulge/disk) stars 
are distributed uniformly around these clouds. Only the light from the old 
stars that are behind the cloud would suffer high extinction, whereas the 
light from the rest of the stars would have a maximum reddening 
of \ebv=0.18 mag. The net extinction
of the old population happens to be 43\% of that of the young population.
Galaxies with a ratio considerably lower than the mean ratio (e.g. NGC\,1022) 
may be those where the young star-forming region is located at the far-side 
of the bulge/disk, and vice versa.

Analyzing a sample of starburst nuclei, Calzetti et al. (1994)
noticed that the continuum near \ha\ and \hb\ suffered about half
the attenuation as suffered by the nebular lines. 
Our result would be identical to that obtained by Calzetti et al.,
if the near ultraviolet and optical continua are predominantly 
contributed by the old stars, i.e. stars that do not supply ionizing photons.
We have demonstrated that this is indeed the case in previous subsections. 
Hence our detailed analysis reinforces the interpretation put forward
by Calzetti et al. (1994) that the reason for a lower continuum extinction 
as compared to the extinction of nebular lines is
the relatively lower extinction towards the 
older stars as compared to extinction towards the young massive stars.
Our analysis also suggests, for instance, that the extinction ratio 
determined from the optical spectrum cannot be simply extrapolated
to the ultraviolet, because of the different contribution of the old and
the young populations at these wavelengths (see e.g. Panuzzo et al. 2003, for
a thorough discussion of this problem).

In most cases, we find the reddening of the intermediate-age population 
to be lower than that of the old population. 
Is it a real effect or is it due to some inherent limitation of the model 
used? In order to address this question, we go back to Figure~3, where the
typical contributions of the three populations are plotted. It can be seen
that the wavelength range where the intermediate-age population contributes,
if at all, covers less than 500~\AA --- between the blue 
limit ($\sim 3600$~\AA) and H$\delta$. 
In this range, there are only two galaxies (NGC\,470 and NGC\,1022), 
in which the intermediate-age population is the dominant contributor.
Significantly, the derived extinctions of the intermediate and old 
populations compare well in these two cases. In the rest of the galaxies,
the available baseline is too small to reliably estimate the reddening.
The amount of intermediate age population in most galaxies is basically
determined from the observed strength of the H$\delta$ line.
As our criterion was to choose models that have minimum extinction (or 
equivalently minimum mass-to-light ratio), the models are able to give only 
a lower limit to the extinction of the intermediate-age population. 
A fair estimate of the mass of the intermediate-age population may be 
obtained by assuming that it has the same reddening as that of the old
population. We find that this mostly affects those cases where the mass of
the intermediate-age population is significantly less than that of the
young population. In such cases, the masses estimated this way turn out 
to be almost equal to those of the young population. In rest of the cases,
the masses increase by around 40\%.

\subsection{Effect of older populations on the Balmer EWs}

The flux ratios and equivalent widths of Balmer emission lines are often
used as  diagnostics of extinction and age of the most recent burst
in starburst regions. The existence of previous generations of stars,
if not taken into account, can seriously affect the determination of
both the extinction and age of the young burst.
Older populations affect in two ways --- 
(a) the intermediate-age populations have strong Balmer absorption lines 
in their spectra, which can decrease the flux of the emission lines,  
(b) old population can contribute significantly to the continuum, 
which can lower the emission equivalent width. The latter effect is
often referred to as dilution.
For an accurate determination of the extinction value it is necessary 
to correct for the contamination from underlying stellar absorption lines,
and for the age determination it is necessary to know the dilution
factor, $f_{\rm dil}$, which is the ratio of the observed continuum 
flux to that of the young burst.
The following equations are used to calculate the true emission line 
flux, $f{\rm (line)}$ and equivalent width, $W_{\rm young}{\rm (line)}$, 
from the observed equivalent width, 
$W_{\rm obs}{\rm (line)}$, and continuum flux, $f_{\rm tot}{\rm (cont)}$: 
\begin{equation}
f{\rm (line)} = (|W_{\rm obs}{\rm (line)}| 
                      + \Delta W{\rm (line)})\times f_{\rm tot}{\rm (cont)}.
\end{equation}

\begin{equation}
W_{\rm young}{\rm (line)} = (|W_{\rm obs}{\rm (line)}| 
                       + \Delta W{\rm (line)})\times f_{\rm dil}. 
\end{equation}
$\Delta W{\rm (line)}$ is the correction term, which McCall et al. (1985) 
found to be 2\,\AA\ for the prominent Balmer lines. McCall et al. 
determined this value empirically for a large number of \hii\ regions, 
by forcing the extinction corrected ratios of the first three Balmer 
lines to their photo-ionization values. On the other hand, 
Veilleux et al. (1995) measured the absorption EWs of H$\beta$ using 
simultaneous fitting of the emission and absorption lines for a 
large sample of infrared luminous galaxies. 
The measured values ranged from 1--3\,\AA, with a median value of 2\,\AA.

The decomposition of the observed spectra into the spectra of individual
populations, allows us to make an independent determination
of $\Delta W{\rm (line)}$.
$\Delta W{\rm (line)}$ depends on the equivalent width and the relative 
continuum contribution of intermediate age and old populations. We summed 
up the spectra of intermediate age and old stellar populations to get a pure 
absorption-line dominated spectrum for each of our starburst nucleus. 
We then measured the EW and continuum fluxes for the first four 
lines of the Balmer series in both the total and absorption-line dominated 
spectrum. $\Delta W{\rm (line)}$ and the dilution factor $f_{\rm dil}$
are then simply given by
\begin{equation}
\Delta W{\rm (line)} = W_{\rm old}{\rm (line)}\times
\frac{f_{\rm old}{\rm (cont)}}{f_{\rm tot}{\rm (cont)}},
\end{equation}
and
\begin{equation}
f_{\rm dil} = \frac{f_{\rm tot}{\rm (cont)}}{1-f_{\rm old}{\rm (cont)}},
\end{equation}
where $W_{\rm old}{\rm (line)}$, and $f_{\rm old}{\rm (cont)}$ are
the equivalent width and the continuum flux of the old population alone.

Table~4 contains the statistical properties of $\Delta W{\rm (line)}$ 
and $f_{\rm dil}$ for the four Balmer lines. The mean underlying absorption
correction to \ha\  is $1.6\pm0.3$\,\AA, whereas for the other three lines, it
is close to $2.4\pm0.8$\,\AA. 
The errors on the EWs are the rms deviations for the sample, and not errors
in the measurements. Hence the difference in the mean values for \ha\
and the rest of the lines is significant, although the average values 
are within the quoted errors.
On the other hand, it is a practice to use a correction of 2\,\AA\  for all 
the lines, following McCall et al. (1985). It may be recalled that McCall 
et al. had forced the correction to be the same for all the Balmer lines. 
In this work, we show that this assumption is not strictly true,
and the average value of $1.9\pm0.3$\,\AA\ obtained by McCall et al. is a
compromise value for all the lines.
We emphasize that in a given galaxy correction to \ha\ is always smaller 
than that for the other three lines. The average value of correction 
could be sample-dependent. This is because
the exact numerical value of the correction 
depends on the relative contribution of the intermediate age population, which
could be different in different star-forming regions.
For instance, Kennicutt (1992) found a mean value as high as
5\,\AA\ for the integrated spectra of galaxies.
We believe the average values obtained by us are representative of starburst
nuclei of nearby galaxies and hence we recommend the use of these values,
for a reliable estimation of A$_{\rm v}$ and metallicity from nebular lines
in starburst nuclei.

The mean dilution factors of the \ha\ and \hb\ EWs are $4.7\pm1.9$ 
and $3.2\pm1.2$, respectively for our sample of starburst nuclei. 
Observed EWs of starburst nuclei are generally lower than those
expected for young bursts (e.g. Veilleux et al. 1995). 
The analysis carried out in our work supports the idea that an
underlying old population is the primary cause for the observed low values
of the Balmer EWs in starburst nuclei.

\begin{deluxetable}{lcccccccc}
\tablewidth{0pc}
\tablecaption{Correction to EWs and dilution factors}
\tablehead{
\colhead{} 
                 & \multicolumn{4}{c}{$\Delta W{\rm (line)}$\,\AA}
 &\multicolumn{4}{c}
{$f_{\rm dil} = \frac{f_{\rm tot}{\rm (cont)}}{1-f_{\rm old}{\rm (cont)}}$}
                  \\
\colhead{}    
                 & \colhead{H$\alpha$}  & \colhead{H$\beta$}
                 & \colhead{H$\gamma$} & \colhead{H$\delta$}
                 & \colhead{H$\alpha$}  & \colhead{H$\beta$}
                 & \colhead{H$\gamma$} & \colhead{H$\delta$}
}
\startdata
Average & 1.57 & 2.48 & 2.32 & 2.49 & 4.7 & 3.2 & 2.7 & 2.4 \\
Rms     & 0.26 & 0.60 & 0.74 & 0.91 & 1.9 & 1.2 & 0.9 & 0.8 \\
Min     & 1.15 & 1.48 & 1.20 & 1.22 & 2.3 & 1.6 & 1.4 & 1.3 \\
Max     & 2.11 & 3.33 & 3.28 & 3.81 & 22.5& 31.3& 40.2& 46.2\\
\hline
\enddata
\end{deluxetable}
\section{Summary and Conclusions}

We have analyzed optical spectra of a sample of starburst galaxies that have
their far infrared, 10$\mu$m, and K-band luminosities similar to those of 
the prototype starburst M\,82. We have adopted a theoretical approach 
that allows us to resolve their star formation history into three major 
episodes, the on-going star formation ($<8$~Myr), the intermediate age 
population (50--500~Myr) and the old populations (older than 1~Gyr).

The extinction of these three main populations is a free parameter and is
fixed, together with the star formation history, by the simultaneous 
fitting of the EWs of the main hydrogen lines and a number of narrow 
bands in the continuum. These data are complemented by the K-band flux, 
estimated for the area within our slit from existing
aperture photometry. The FIR emission could also be used to constrain the
model, but its uncertain contribution within our slit prevented 
us to use it as a firm constraint.

All galaxies have clear signatures of the above three main episodes of star
formation, with birthrate parameter (current SFR over average past value), 
between 1 and 12. The mass of the intermediate population is generally 
comparable to the mass formed during the current burst and, together, they 
provide only a small contribution to the total mass sampled by our spectra.
The spectra sample approximately the central 1 kpc region, and the mass 
within this region turns out to be between 10$^9$\msun\  and 
2$\times$10$^{10}$\msun, with an average value 
$\simeq$8$\times$10$^9$\msun. These values are within the mass range
found for bulges of spiral galaxies.
There is a clear tendency for the most massive bulges to host a stronger
starburst activity. The mass of the young component increases linearly 
with that of the old bulge: M$_{\rm Young}$$\simeq$0.003M$_{\rm Old}$. 

The relation between the mass of the young and old component is the 
most likely cause of the correlation reported by Devereux (1989) between 
the 10$\mu$m luminosity, which is proportional to the present SFR 
(i.e. the mas of the young component) and the K-band luminosity, 
which is proportional to 
the old component. This correlation led Devereux to conclude that the K-band 
luminosity is mainly produced by red supergiant stars of the young component. 
On the contrary, our population synthesis technique allows us to exclude 
this possibility. In the majority of the cases, the old population contributes 
more than 60\% of the flux to the K-band, and along with the intermediate-age
population it contributes more than 40\%  even to the blue continuum.
Test cases show that it is not possible
to reproduce the broad features of the continuum and in particular the depth of
the \caiik\  line by suppressing the older population, even by assuming a
higher metal content. Hence, in most of the M\,82-like starburst galaxies, 
the K-band luminosity can be used as an estimate of stellar mass.
However, when the birthrate parameter exceeds the average value of 
our sample galaxies ($\simeq$5), the young populations
start contributing a significant fraction to the NIR luminosity.
This should be taken into account when trying to determine, from the
restframe optical-NIR luminosity, the mass of the old component in high 
redshift galaxies, where at a given SFR the birthrate parameter is likely
to be larger.

The regions with on-going star formation have systematically higher extinction
than the older populations: A$_{\rm v}$(old)/A$_{\rm v}$(young)$=0.43\pm0.13$.
The fact that the continuum of starburst nuclei is mainly contributed by the
old population, makes this result identical to that obtained by 
Calzetti et al. (1994), who found A$_{\rm v}$(continuum)/A$_{\rm v}$(emission 
line)=0.50. It was not possible to reliably determine the extinction of 
the intermediate-age population due to the small base wavelength over 
which this population dominates, if at all, in the sample nuclei.

The decomposition of the observed spectra into spectra of individual 
populations, allowed us to determine the underlying absorption EWs 
for Balmer lines. 
The resulting mean value for \ha\ is $1.6\pm0.3$\,\AA, and
around $2.4\pm0.8$\,\AA\ for the rest of the Balmer lines, where the
errors represent the real variations from galaxy to galaxy and are not
due to measurement errors. 
Hence in a given galaxy correction to \ha\ is always smaller
than that for the other three lines.
These values are improvements over the values suggested by McCall et al. 
(1985), who obtained the correction by forcing it to be the same for 
the first three Balmer lines. As expected, the value of $1.9\pm0.3$\,\AA\ 
that was obtained by McCall et al. is intermediate between that of \ha\ 
and the rest of the Balmer lines.
The continuum from the old population plays an important 
role in reducing (or diluting) the observed emission EWs below the
values expected for a purely young burst. We find that the mean dilution 
factors for \ha\ and \hb\ are $4.7\pm1.9$ and $3.2\pm1.2$, respectively.

\acknowledgments

We thank Bianca Poggianti, Pasquale Panuzzo, Alberto Franceschini and 
Roberto Terlevich for thoughtful discussions. 
We also thank the referee for several useful comments, which
have enabled us to improve the original manuscript.
This work was partly supported by the CONACyT 
projects 39714-F, 36547-E and J37680-E. A. B. thanks INAOE for its 
kind hospitality during his visits. 
This research has made use of the NASA/IPAC Extragalactic Database, 
which is operated by the Jet Propulsion Laboratory, California
Institute of Technology, under contract with the National Aeronautics and Space
Administration. 


\end{document}